\documentclass[reprint,amsmath,aps,prb,footinbib,notitlepage,longbibliography,superscriptaddress,floatfix]{revtex4-2}

\usepackage{epsfig}
\usepackage{graphicx}
\usepackage{amsmath}
\usepackage{mathtools}
\usepackage{amscd}
\usepackage{amssymb}
\usepackage{amsfonts}
\usepackage{amsthm}
\usepackage{enumitem}
\usepackage{stmaryrd}
\usepackage{tikz-cd}

\usepackage[T1]{fontenc}

\usepackage{xcolor}

\def\be{\begin{equation}}
\def\ee{\end{equation}}
\def\bea{\begin{eqnarray}}
\def\eea{\end{eqnarray}}

\def\bfk{\mathbf{k}}

\newcommand{\bra}[1]{\left\langle #1 \right|}
\newcommand{\ket}[1]{\left|#1\right\rangle}
\newcommand{\sbra}[1]{\langle #1 |}
\newcommand{\sket}[1]{|#1\rangle}
\newcommand{\braket}[2]{\left\langle#1 |  #2\right\rangle}

\newcommand{\sbraket}[2]{\langle#1 |  #2 \rangle}

\newcommand{\squish}[3]{\left\langle#1 |  #2 | #3 \right\rangle}

\newcommand{\sex}[1]{\langle #1 \rangle}

\usepackage{physics,siunitx,hyperref,bm}

\begin{document}

\title{Low-energy tail of the spectral density for a particle interacting with a  quantum phonon bath}

\author{Donghwan Kim}
\affiliation{Department of Chemistry and Chemical Biology, Harvard University, Cambridge,
Massachusetts 02138, USA}

\author{Bertrand I. Halperin}
\email{halperin@g.harvard.edu}
\affiliation{Department of Physics, Harvard University, Cambridge, Massachusetts 02138, USA}

\date{\today}

\begin{abstract}

We describe two approximation methods designed to capture the leading behavior of the low-energy tail of the momentum-dependent spectral density $A(\bfk, E)$ and the tunneling density of states $D(E)$ for an injected  particle, such as an electron or an exciton, interacting with a bath of phonons at a non-zero  initial temperature  $T$, including quantum corrections due to the non-zero frequencies of the relevant phonons. In our imaginary-time-dependent Hartree (ITDH)  approximation, we consider a situation where the particle is injected into a specified coherent state of the phonon system, and we show how one can use the ITDH approximation to obtain the correlation function $C(\tau)$ for that  initial state. The thermal average $C(\tau)$ is  obtained, in principle, by integrating the result over all possible initial phonon coherent states, weighted by a thermal distribution. However, in the low-energy tail, one can obtain a good first approximation by considering only initial states near the one that maximizes the integrand.  Our second approximation, the fixed-wave-function (FWF) approximation,  assumes that the wave function of the injected particle evolves instantaneously to a wave function which then is independent of time, while the phonon system continues to evolve due to interaction with the particle. 
We discuss how to invert the Laplace transform and how to obtain $A(\bfk,E)$ as well as $D(E)$ from the imaginary-time analysis.  The FWF approximation is used to calculate $D(E)$ for a one-dimensional continuum model of a particle interacting with acoustic phonons, and effects due to the quantum motion of phonons are observed.  In the classical phonon limit, where the nuclear mass is taken to infinity while the elastic constants and other parameters are held fixed, the dominant behaviors of both the ITDH and FWF approximations in the low-energy tail reduce to that found in the past for a particle in a random potential with  a Gaussian statistical distribution.

\end{abstract}

\maketitle

%\widetext

\section{Introduction}

In 1957, Rashba and Davidov published an article in Ukrainian in  the Ukrainian Physical Journal on optical absorption in a molecular crystal with a weak interaction between excitons and phonons \cite{DavidovRashba57}.  Also in 1957,  Rashba published a pair of  articles in the Russian journal Optika i Spektroskopiya on the theory of electronic excitations interacting with lattice vibrations in a molecular crystal \cite{Rashba57a,Rashba57b}.   These articles  considered cases  of  both light and heavy excitons, distinguishing excitons whose band width in the absence of lattice distortions is  large or small on a scale depending on the strength of the particle-phonon interactions, with particular attention to the case of strong interactions. The three papers considered one-dimensional chains as well as three-dimensional crystals. 

An important question, addressed in Ref. \cite {DavidovRashba57}, was how do phonons affect the line shape for optical absorption associated with the creation of an exciton, at finite temperatures as well as at $T=0$?  
In simple cases, the optical absorption for a photon of frequency $\omega$ will be proportional to the momentum-resolved spectral density $A(\bfk,E)$ at energy $E=\hbar \omega$  for an exciton injected  with momentum $\bfk =0$.  A related quantity is the tunneling density of states  $D(E)$ for a particle such as an exciton or an electron  injected at a single point, which is equal to the integral of $A(\bfk,E)$ over all values of the momentum $\bfk$.

Reference \cite{DavidovRashba57} and many subsequent works have  examined  the absorption line shape near the peak of the spectrum, where the absorption is relatively large, or in the high-energy tail, where the absorption falls off as an inverse power of the energy. 
 In general, these regions of the spectrum can be understood by using perturbation theory or related diagrammatic methods to treat exciton-phonon interaction.   In the present paper, however, we shall be concerned with the low-energy tail of the spectrum, where the density of states is very small and the absorption is very weak. 
This is a region where methods based on perturbation theory are generally inadequate and other approaches must be used. 

In many insulating materials, particularly alkali halides and other materials with a relatively large band gap and tightly bound excitons, the optical-absorption coefficient at photon energy $E$, for a range of  temperatures $T$, has been fit by the empirical Urbach formula:
\be
\label{urbach} 
\alpha (E) = \alpha_0 e^{- \sigma (E_0 - E) / T}, 
\ee
where $\alpha_0, \sigma, E_0$ are material parameters, with, typically, $\sigma \approx 1$.
(See, e.g., \cite{Knox63,DowRedfield72,Toyozawa3} and references therein. We use units where $k_B = \hbar = 1$.) 

Attempts to explain (\ref{urbach}) have generally treated the phonon bath as giving rise to lattice distortions that can be treated as static on the time-scale of interest for the absorption process. In this case, the problem reduces to the model of a particle interacting with a potential that obeys Gaussian statistics with variance that will be  proportional to $T$, for $T \gg \omega_{\rm{ph}} / 2$, where $\omega_{\rm{ph}}$ is a characteristic phonon frequency \cite{Toyozawa64,HL1,Kim22}.
The low-energy tail is then produced by processes in which  the exciton is injected in a region where a thermal distortion has led to a reduction in the local energy gap.  

For $T \leq \omega_{\rm{ph}} / 2$, it is clear that  quantum motion  of the lattice should be taken into account, and one might expect significant corrections to the classical phonon picture. 
Indeed, observations do not necessarily conform to Eq. (\ref{urbach}) at low temperatures.  In some cases,  the law has been seen to apply with the replacement of $T$ by an effective temperature $T^*$ that is of the order of $\omega_{\rm{ph}} / 2$ at low temperatures. However, it is not clear whether the observed low-temperature behavior is controlled by phonons or whether the effects of impurities must be taken into account. 

Explanations of Urbach's rule have been only partially successful,  even at high temperatures where the classical phonon approximation is presumably valid.  Typically, theoretical analyses predict an absorption tail of the form
 \be
\label{modurbach} 
\alpha (E) \sim  \alpha_1   e^{h(E) / T}, 
\ee
where the function $h$ is independent of $T$, and $\alpha_1$ varies relatively slowly with $E$ and $T$ \cite{HL1,HL2,HalperinThesis}.  However,  $h$ is not a perfectly linear function of  $E$ in any obvious model, and there is no clear  reason why $dh/dE$ should turn out to be very close to 1.  Nevertheless, numerical calculations have  produced absorption curves that are in rather good agreements with (\ref{urbach}) in at least some instances \cite{DowRedfield72,Toyozawa3}.

In the present paper, we shall not attempt to provide further insight into the remarkable validity of (\ref{urbach}) in the high-temperature regime.  Rather, we wish to explore methods for including the dynamic effects of lattice vibrations at temperatures where they might become important.  We introduce  two related approximations, which we expect will describe  the most important dependences of the spectral density and tunneling density of states, (on a logarithmic scale), in the region of the low-energy tail, in a system without impurities. We note that impurities may well be important in many cases, and indeed there have been many theoretical investigations of low-energy behavior of $D(E)$ and $A(\bfk,E)$ for particles in a random potential due to impurities \cite{HL1,HL2,FrischLloyd,Lifshitz,ZittartzLanger,Baranovskii,Mieghem92}.  Nevertheless, we believe that it is important at least as a matter of principle to understand the effects of phonons in an ideal system.    

The effects of phonon interactions on the behavior of an isolated electron or exciton have been studied extensively over the years in the context of the polaron problem.  These investigations have largely concerned such questions as the phenomenon of self-trapping in the presencece of strong particle-phonon interactions, and the  binding energy and effective mass of the resulting polaron, as well as polaron mobility at low temperatures, rather than the spectral properties of interest to us here (cf. \cite{Giustino17,franchini2021polarons,devreese2020frohlich} and references therein). The concept of self-trapping will play a role in the discussion below, however.   

Our principal approach to the low-energy tail problem makes use of correlation functions in imaginary time, of the particle creation and annihilation operators, which are Laplace transforms of the density of states and spectral densities in which we are interested. After defining our model in the following section, we discuss a procedure for inverting the Laplace transform that is applicable to the low-energy behavior  of the functions we wish to calculate. In Sec. \ref{ITDH} we describe our principal approximation, which we denote as the imaginary-time-dependent Hartree (ITDH) approximation.  In Sec. \ref{FWF} we describe a fixed-wave-function (FWF) approximation, which we expect to be less accurate than the ITDH approximation but easier to apply. In Sec. \ref{classical}, we discuss the behavior of these approximations in the classical phonon limit, where we find that the two approximations are essentially equivalent and are closely related to previous studies of the low-energy tail in a Gaussian random potential.  The FWF approximation is  applied in Sec. \ref{continuum} to a one-dimensional model of a particle interacting with quantum-mechanical acoustic phonons in the continuum limit. 
While the bulk of our paper is focused on obtaining an optimum estimate of the tunneling density of states $D(E)$, the analysis is easily extended to predict the momentum-dependent  spectral density $A(\bfk,E)$. This is discussed in Sec. \ref{AkE} .  

Although  our paper is presented largely in the context of one-dimensional models, the methods should be useful,  with some possible modifications, for three-dimensional problems.  These are discussed in Sec. \ref{3D}. 

In Appendix \ref{EOMITDH}, we present  a derivation of the imaginary-time-dependent Hartree equations of motion.
Some higher-order corrections, which contribute pre-exponential corrections to the density of states in the classical phonon limit, are discussed in Appendix \ref{prefactors}.   Other appendixes discuss  details of the density of states maximization related to a modified form of the FWF, statistics of the potential fluctuations in a thermal ensemble with and without quantum corrections,  and an additional method for obtaining $D(E)$  in the FWF approach. 

Our principal results are summarized in Sec. \ref {conclusions}.

\section{Model}
\label{model}

We consider here a one-dimensional model of particles and phonons on a lattice.  We assume  lattice constant $a$, with periodic boundary conditions and $N \gg 1$ sites, giving total length $L=Na$.  We assume a Hamiltonian of the general form
\be
H = H_e + H_p + H_{ep} ,
\label{generalH}
\ee
\be
H_e = \sum_k  \epsilon_k c^\dagger_k c_k ,
\ee
\be
H_p =  \sum_k \omega_k a^\dagger_k a_k , 
\ee
\be
H_{ep} = a \sum_k \sum_x \psi^\dagger_x \psi_x (  \lambda_k a_k^\dagger + \lambda_{-k}^* a_{-k} ) e^{-ikx} ,
\label{Hep}
\ee
where $\epsilon_k$ and $\omega_k$ are dispersion relations for particles and phonons, respectively, $c_k$ ($a_k$) and $c_k^\dag$ ($a_k^\dag$) are annihilation and creation operators for a particle (a phonon) with momentum $k$, while  $\lambda_k$ is the coupling strength for interaction between a particle and a phonon with momentum $k$, and $\psi^{\dagger}_x$ is a creation operator for a particle on a lattice site at position $x = n a$, with $n$ an integer:
\be
 \psi^{\dagger}_x = L^{-1/2} \sum_k e^{-ikx} c^{\dagger}_k .
 \ee
We are using a normalization such that
\be
\comm{\psi_x}{\psi^{\dagger}_{x'}}_{\mp}
\equiv
\psi_x\psi^{\dagger}_{x'}
\mp \psi^{\dagger}_{x'}\psi_x
= a^{-1} \delta _{x x'},
\label{commRel}
\ee
with $-$ for a boson and $+$ for a fermion, which will facilitate passing to the continuum limit, replacing $a \sum_x$ by $\int dx$.
We shall assume the Hamiltonian is time-reversal invariant, so $\lambda_{-k}=\lambda_k^*$.

It will be helpful to rewrite the coupling constants $\lambda_k$ as
\be
\lambda_k = (2N M \omega_k )^{-1/2}  \gamma_k ,
\ee
where $M$ is the nuclear mass.  Then $\gamma_k$ will   remain constant if we vary  $M$ while keeping fixed the elastic constants $M \omega_k^2$  and keeping fixed the deformation potential felt by a particle for a given displacement of the atoms. The classical phonon model can then be described by taking $M$ to infinity and $\omega_k \to 0$, keeping $\gamma_k$ fixed. The coefficients $\lambda_k$ have dimensions of energy, while  $\gamma_k$ has dimensions of energy per unit length. 

We wish to calculate the low-energy tail of the particle density of states $D(E)$.  Specifically, we consider an initial state described by a thermal distribution of phonons, with no particle present,   described by the initial density matrix
\be
w = Z^{-1}  e^{-H_p / T} ,
\ee
where $Z = {\rm{tr}}\ e^{-H_p / T}$,
and we wish to calculate the distribution of possible energy changes produced by the added particle. If we work in the basis of exact eigenstates of $H$, then we may write
\be
\label{rhodef} 
D(E) = Z^{-1} \sum_f \sum_i | \squish {f} {\psi_{x_0} ^\dagger} {i} |^2 \delta (E - E_f + E_i)  e^{-E_i / T} ,
\ee
where $x_0$ is an arbitrary point on the lattice, and the sum is over eigenstates of the Hamiltonian with no particle present for the state $i$ with energy $E_i$ and one particle present for the state $j$ with energy $E_j$.   The result will be independent of the choice of $x_0$, as the model is translationally invariant. 

Alternatively we may  calculate the correlation function
\be
\tilde{C} (t) =   {\rm {tr}}  \, (e^{i H t}  \psi_{x_0}  e^{- i H t} \psi^\dagger_{x_0}  \, w)  = \int_{- \infty} ^{\infty}  e^{-i t E} D(E) dE ,
\label{corrFourier}
\ee
or its imaginary-time version
\begin{eqnarray}
\label{laplace}
C(\tau) \equiv  \tilde {C} (-i \tau) & = &   {\rm {tr}}  \, (e^{ H \tau}  \psi_{x_0}  e^{-  H \tau } \psi^\dagger_{x_0} \, w) \nonumber \\
  & = &  \int_{- \infty} ^{\infty} e^{- \tau E} D(E) dE.
\end{eqnarray}
In principle, if $C(\tau)$ is accurately known, it can be analytically continued to obtain $ \tilde C$, which can then be Fourier transformed to obtain $D(E)$. However, an approximate solution at real times would not be useful for obtaining the low-energy tail because the Fourier transform would be very unstable to small errors due to the oscillating nature of the integrand.
In the following section,  we argue that knowing $C(\tau)$  we can in fact directly obtain a good estimate of the low-energy tail of $D(E)$ under appropriate circumstances.

The momentum-dependent spectral density $A(k,E)$ is defined by replacing the operator $\psi^\dagger_{x_0}$ in Eq. (\ref{rhodef}) with the operator $c^\dagger_k$:
\begin{align*}
    A(k,E)=Z^{-1}\sum_{i,f}\abs{\mel{f}{c_k^\dag}{i}}^2\delta(E-E_f+E_i)e^{-E_i/T}.
\end{align*}
Similarly we may define a momentum-dependent correlation function $C(k,\tau)$ by replacing $\psi_{x_0}$ and $\psi^\dagger_{x_0}$ in (\ref{laplace}) by $c_k$ and $c^\dagger_k$. The function $C(k,\tau)$ will be the Laplace transform of $A(k,E)$: 
\begin{eqnarray*}
C(k,\tau) =  {\rm {tr}}  \, (e^{ H \tau}  c_k  e^{-  H \tau } c_k^\dag \, w) 
 =  \int_{- \infty} ^{\infty} e^{- \tau E} A(k,E) dE.
\end{eqnarray*}

\section{Inverting the Laplace Transform} 
\label{invert}

Let us write  $C(\tau)$ and $D(E)$ in the form
\be
C(\tau) = e^{g(\tau) } ,  \,\,\,\,\,\, D(E) = E_0^{-1} e ^ { f(E) } ,
\ee
where $E_0$ is a characteristic energy, such as the width of the peak of the spectral density $A(k,E)$ at $k=0$.  The low-energy tail is a region of energies less than some energy $E_1$ that is of order $E_0$ below the nominal bottom of the band where we have $f(E) < -1$ [such that $D(E)$ is small enough], and $f'(E) > E_0^{-1}$ [such that $D(E)$ decreases rapidly with decreasing  $E$].  We shall make the additional assumption here that $f''(E) < 0 $  for all $E< E_1$, which, following a Halperin--Lax-type analysis, should be a good assumption for d=1 and marginally valid for d=2, but generally false for d=3. (Possible modifications to handle the case of d=3 will be discussed in Sec. \ref{3D}.) 

Under these circumstances, Laplace's method can be used to evaluate the integral (\ref {laplace}); the integral should be dominated by the region near the maximum of the integrand, where $E$ takes on the value $E_\tau$, for the given $\tau$, such that 
\be 
\label{taufprime}
f ' (E_\tau) = \tau .
\ee
In the neighborhood of $E_\tau$, we can expand $f$ as
\be
f (E_\tau+\delta E) = f(E_\tau) + \tau \delta E + (\delta E)^2 (d \tau / d E_\tau) /2  + ...  \, ,
\label{exponentExpansion}
\ee
with $d\tau / d E_\tau = f'' (E_\tau) < 0 $. 
If we ignore terms  higher-order in $\delta E$, we may evaluate  the integral in  (\ref{laplace}) with the result 
\be
C(\tau) \approx D(E_\tau ) e^{- \tau E_\tau} \left| 2 \pi \frac {d E_\tau} {d \tau} \right| ^ {1/2} .
\label{CQuadAprx}
\ee
Defining $\tau_E$ by the equation $\tau_E= f ' (E)$, we may invert the above equation to give
\be 
D(E) \approx    ( 2 \pi )^{-1/2} C(\tau_E) e^{  \tau_E  E } \left| \frac {d \tau_E}{d E} \right| ^{1/2 } .
\ee

In the low-energy tail, we expect that the last factor in the above equation should have a weaker dependence on $E$ than the earlier factors, because $\tau$ depends linearly on $f'$, whereas the other factors depend exponentially on $f$.   Ignoring  the last factor,  we  see
that if the function $C(\tau)$ is known,  then  $\tau_E$ can be obtained from the requirement that at $\tau = \tau_E$, 
\be
\label{EdlnCdt}
E = -  \frac { d \ln C}{d \tau}    .
\ee
Also, 
\be
\frac {d \tau_E}{dE} = - \left( \frac {d^2 \ln C }{d \tau^2} \right) ^ {-1} .
\ee

If the zero of energy is chosen at the bottom of the unperturbed electron energy band, then $E$ will be negative in the low-energy tail, and $dC/d \tau$ will be positive. Moreover, large negative values of $E$ will correspond to large positive values of $\tau$.

 The above arguments can be made more precise if one can define a small parameter $\zeta$, such that in the limit $\zeta \to 0$, with $E$ held fixed,
 \be
 \label{zeta}
 \zeta \, \ln D(E) \to \tilde{f}(E),  
 \ee
 where $\tilde{f}(E)$ is independent of $\zeta$. Laplace's method becomes exact in the limit $\zeta \to 0$.

 For the systems we consider here, we find that for $E$ lower than  the minimum of the unperturbed spectrum $\epsilon_k$, the function  $D(E)$ can be written a form similar to (\ref{zeta}) in the limit where $T$ and $M^{-1}$ are small,  while material parameters such as $\gamma_k, \epsilon_k,$ and $M \omega_k^2$ are held fixed.  Note that the phonon frequencies will be proportional to $M^{-1/2}$.  In the case where $T$ is comparable to or larger than the typical frequency $\omega_{\rm{ph}}$ of the phonons most important for states in the low-energy tail, the parameter $\zeta$ will scale proportional to $\tilde{\gamma}^2 T$, where $\tilde{\gamma}$ is a typical value of the coupling constant $\gamma_k$.  In the limit where $T\to 0$, with   $\omega_{\rm{ph}}$    small but finite, we find that $\zeta \propto \tilde{\gamma}^2 \omega_{\rm{ph}}$, provided that $E$ is larger than $E_{\rm{min}}$, the ground state energy for a single particle coupled to the phonons. Quantum corrections will be most important when $T\lesssim\omega_{\rm{ph}}$.
 
 Of course, the actual value of $T$ or $\omega_{\rm{ph}}$ necessary to be in the low-energy tail will depend on details of the system, including the energy in question.  In experiments, one may enter the low-energy tail region by varying the measurement energy rather than the temperature. 
 
 The approximations employed in this paper are intended to give a good approximation to the function $\tilde{f}(E)$,  which means that they should give a good approximation to the leading exponential behavior of $D(E)$ and $A(\bfk,E)$ in this limit. However, 
 they are not expected to give correctly the pre-exponential factors.

An important caveat is that in some situations, we may encounter examples where $f''(E) > 0$ for an energy range of interest, $E_b < E < E_a$. In such cases the Laplace transform cannot be inverted in this simple way. However, we should  get a warning of this problem from studying $C(\tau)$ because 
 as $\tau$ is varied the maximum of the integrand in (\ref{laplace}) will jump rapidly between an energy $E_c > E_a$ to an energy $E_d < E_b$, so that  Eq. (\ref{EdlnCdt})  will move rapidly through  the  energy range of interest.  
 We shall return to this issue in Sec. \ref{3D}, but at least in the one-dimensional models we focus on here, this problem will not arise.

\section{imaginary-time-dependent Hartree (ITDH) Approximation  }
\label{ITDH}  

\subsection {Coherent state representation} 

Let $\sket{ \{\alpha_k\} }=\ket{\bm\alpha} $ denote a coherent phonon state with $\sex {a_k} = \alpha_k$, and the value of $\alpha_k$ is specified for every wave vector $k$.  The initial density matrix $w$ may be written as
\be
w =      \int d \bm\alpha \,   w_{ \bm\alpha }  \,   \sket { \bm\alpha }  \sbra{ \bm\alpha } ,
\ee
where the integral is over the real and imaginary parts of the variables in $\bm\alpha$, i.e., $d\bm\alpha=\prod_k[d(\Re\alpha_k) d(\Im\alpha_k)]$,
\be
w_{\bm\alpha} = Z^{-1}_{\rm{coh}}  \exp [- \sum_k (  | \alpha_k |^2 /  n_k) ] ,
\label{walpha}
\ee
\be
 n_k
 \equiv
 \frac {1} { e^{ \omega_k / T} -1}    ,
\ee 
and  
\be
Z_{\rm{coh}} = \prod_k (\pi n_k).
\ee
We may now write \eqref{laplace} as
\begin{align}
    C(\tau)=\int d\bm\alpha w_{\bm\alpha}C_{\bm\alpha}(\tau)
    \label{therm}
\end{align}
where
\begin{align}
    C_{\bm\alpha}(\tau)=\bra{\bm\alpha}e^{H\tau}\psi_{x_0}e^{-H\tau}\psi_{x_0}^\dag\ket{\bm\alpha}.
\end{align}

\subsection  {Factorization approximation} 

We next  make the factorization approximation
\be
e^{-  H \tau} \psi^ {\dagger}_{x_0}  \ket{ \bm\alpha } \approx  a^{-1/2}R(\tau) \Phi_\tau^\dagger \ket{\bm\beta(\tau)}  \equiv \sket {\Psi(\tau)},
\ee
where $  \ket{\{\beta_k(\tau)\}}=\ket{\bm\beta(\tau)}$ is a (normalized) coherent phonon state, 
 $  \Phi_\tau^\dagger $ creates a particle in a state with a normalized wave function $\phi (x,\tau) $:  
\be
\Phi_\tau ^\dagger =  a \sum_x  \phi (x,\tau) \psi^\dagger_x ,
\ee
and $R( \tau)$ is a renormalization factor, chosen as a positive real number depending on $\tau$, made necessary by the imaginary time propagation.  
[Our normalization convention is such  that $a \sum_x |\phi(x)|^2\equiv  \braket {\phi}{\phi} =1$].

At time $\tau=0$, we set $R=1$, $\phi (x) = a^{-1/2}\delta_{x x_0}$, and $\beta_k = \alpha_k$.
The parameters should then  evolve in time according to the equations of motion, which are derived in Appendix \ref{EOMITDH}:
\be
\label{dphidt}
\partial \phi / \partial \tau = - [H_e + V (x,\tau) - E_\phi(\tau) -iQ(\tau)] \phi
\ee 
\be
\label{V}
V(x,\tau) = \sum_k     (  \lambda_k \beta^*_k + \lambda_{-k}^*\beta_{-k}) e^{-ikx} ,
\ee
\be
\label{Ephi}
E_\phi(\tau) = \squish {\phi(\tau)} {H_e + V(x,\tau) } {\phi(\tau)} ,
\ee
\be
Q(\tau) =  \Im \left[ \sum_k \beta^*_k(\tau) \lambda_k |\phi^2|_k \right]
\ee 
where 
\be 
| \phi^2 | _k  \equiv a \sum_x  | \phi (x, \tau) |^2 e ^{-ikx} ,
\ee
and
\be
d \beta_k / d\tau  = - \omega_ k \beta_k - \lambda_k  |\phi^2| _ k ,
\label{EOMbeta}
\ee
\be
\label{dRdt}
dR / d\tau =  - R \left( E_ \phi + \sum  _k \omega_k   |\beta_k|^2 \right) .
\ee
Without the particle, one obtains
\begin{align}
    e^{-H\tau}\ket{\bm\alpha}=R_0(\tau)\ket{\bm\beta_0(\tau)},
\end{align}
where parameters evolve in time according to the equations of motion
\begin{align}
    d\beta_{0k}/d\tau&=-\omega_k\beta_{0k},
    \\
    dR_0/d\tau&=-R_0\sum_k\omega_k\abs{\beta_{0k}}^2,
    \label{dR0dtau}
\end{align}
whose solutions are
\begin{align}
    \beta_{0k}(\tau)&=\alpha_k e^{-\omega_k\tau}
    \\
    R_0(\tau)
    &=
    e^{-\sum_k\abs{\alpha_k}^2(1-e^{-2\omega_k \tau})/2}.\label{drdt}
\end{align}
Thus, we find
\be
\label{Calphatau}
C_{\bm\alpha} (\tau) \approx a^{-1/2}  \frac{R(\tau)}{ R _0(\tau)}  \phi(x_0, \tau)  \prod_k \sbraket {\beta_{0k}(\tau)} {\beta_k(\tau)} ,
\ee
where for complex numbers $z_1$ and $z_2$, the inner product of the corresponding coherent states is
\be
\sbraket {z_1}{z_2} = \exp\{z_1^*z_2-|z_1|^2/2 - |z_2| ^2 /2 \} .
\ee

Now, in principle, we should solve equations of motion for all possible choices of the initial variables $\bm\alpha$ and carry out the integration (\ref{therm}). 
However, in order to get a binding energy in the low-energy tail, one needs an initial phonon configuration with a large distortion, costing an energy large compared to $T$. The density of states will then be dominated by the configuration that achieves the binding with minimum energy cost. [E.g. for the one-dimensional continuum model considered in \cite{HL1}, in the classical phonon limit, the optimum distortion is proportional to $\sech^2\kappa(x-x_0)$, with $\kappa=(2m\abs{E})^{1/2}$.] Relatively small departures that still preserve the binding energy  will cost energies larger than $T$, so their probability will diminish rapidly. Correspondingly, we expect the integral \eqref{therm}   for the Laplace transform will be dominated by a region near the optimal phonon configuration where the product $w_{\bm\alpha}C_{\bm\alpha}(\tau)$ is a maximum. 
Thus, using Laplace's method, we may make the further approximation 
\be
\label{CCmax}
C(\tau) \approx S(\tau) C_{max} (\tau),
\ee
where 
\be
C_{max} (\tau) = {\rm{max}}_ {\bm\alpha}   \left[ Z_{\rm{coh}} w_{ \bm\alpha }    \abs{C_{\bm\alpha} (\tau)} \right]  ,
\ee
and the prefactor $S(\tau$) should be evaluated using a perturbative expansion about the maximizing phonon configuration. Essentially, the factor $S$ represents something like the difference in  the entropies of initial and final states.
(We assume that there is a single $\bm\alpha$ that maximizes $w_{\bm\alpha}\abs{C_{\bm\alpha}(\tau)}$ and the corresponding $C_{\bm\alpha}(\tau)$ is real. See Appendix \ref{EOMITDH}.)
In practice, it may be a good first approximation to set $S=1$.  Finally, we should use the procedures of Sec. \ref{invert} to invert the Laplace transform and  calculate $D(E)$.

 \section {Fixed-Wave-Function (FWF) Approximation}
 \label{FWF}
 
 We may simplify the analysis by ignoring the equation of motion (\ref {dphidt}) for the wave function $\phi (x,\tau)$ and simply take $\phi$ to be a time-independent trial wave function $\phi_{\rm{tr}} (x)$, which will depend on the energy $E$ of interest, and  which we will eventually choose to optimize our estimate of $D(E)$.  We retain Eqs. (\ref{V})--(\ref{drdt}) for the time evolution of $\beta_k $ and $R(\tau)$; however, we employ the initial condition 
 \be
 R(\tau = 0) = \phi_{\rm{tr}} (x_0) 
 \ee
 rather than $R(0) = 1$.  
 
 For a given choice of $\phi_{\rm{tr}}$, the problem now reduces to the Franck-Condon problem of a localized electronic excitation linearly coupled to a phonon bath, which was studied, for example, by Lax and Hopfield \cite{Lax52,Hopfield62}.  In particular, using the solution of Lax with $t= -i\tau$, we find
 \be
 \label{CFC}
 C_{\phi_{\rm{tr}}} (\tau) \approx |\phi_{\rm{tr}} (x_0 )|^2  e^ {- \tau E^\phi_e}   e^{F(\tau)} ,
\ee
\be
E^\phi_e = \squish {\phi_{\rm{tr}}}{ H_e}{\phi_{\rm{tr}}} , 
\ee
\be
\label{Fdef}
F(\tau) =  \sum_k \frac {|C_k|^2 }{ \omega_k^2 } [ (n_k +1) e^{-\omega_k \tau} 
+ n_k e^{\omega_k \tau} + \omega _k \tau -(2 n_k +1) ]  ,
\ee
\be
C_k = \lambda_k  |\phi^2 |_k .
\label{eq:Ck}
\ee

Now, we
approximate $C(\tau)$ by choosing the trial function $\phi_{\rm{tr}}$ so as to maximize the value of $C_{\phi_{\rm{tr}}} (\tau)$, as given by (\ref{CFC}) and (\ref{Fdef}), for the given $\tau$:
\begin{align}
    C(\tau)\approx{\rm{max}}_{\phi_{tr}}[C_{\phi_{tr}}(\tau)].
    \label{Cmaximization}
\end{align}
After repeating this for a suitable range of values of $\tau$, one can then invert the Laplace transform using the methods we have described.  

Presumably the FWF estimate  will be less accurate than the ITDH approximation, defined in the previous section. However, it should be easier to compute, and  the two approximations  may not differ much in practice.

\section {Classical Phonon Limit}
\label{classical}

As mentioned above, in the limit where the masses of the nuclei are taken to infinity, while the elastic constants and the temperature are held fixed, the problem we are considering reduces to a calculation  of the density of states of a particle moving in a static random potential $V(x)= \sum_k     (  \lambda_k \alpha^*_k + \lambda_{-k}^*\alpha_{-k}) e^{-ikx}$.

At temperature $T$, the potential obeys Gaussian statistics, with $\sex{V(x)}_{T} = 0$ and a correlation function 
\be
\sex {V(x) V(x') }_{T} = W(x-x'),
\ee
where $\ev{\cdot}_T$ is the thermal average and [c.f. Eq. \eqref{ACcl}]
\be
\label{Wclass}
W(x-x') =T \int_{-\pi/a}^{\pi/a} \frac{dk}{2\pi} \frac{\abs{\gamma_k}^2}{\rho \omega_k^2}
     e^{-ik(x-x')} .
\ee
If the function $V(x)$ is specified, the combinations $\alpha_k + \alpha^*_{-k}$ are thereby determined, but no information is gained about the quantities  $ \eta_k = \alpha_k - \alpha^*_{-k}$. If we integrate $w_{\bm{\alpha}} $ over the variables $\eta_k$, we obtain a probability distribution for the function $V(x)$ (see Appendix \ref{qcweight}): 

\be
w_{\rm{cl}} = Z_{\rm{cl}}^{-1} \exp \left[ -  \frac{a^2}{2}\sum_{x x'} V(x) G(x-x') V(x') \right] ,
\label{wcl}
\ee
where
\begin{align}
    Z_{\rm{cl}}=\int \mathcal{D}V \exp \left[ -  \frac{a^2}{2}\sum_{x x'} V(x) G(x-x') V(x')\right]
\end{align}
is a normalization constant and  the integral is over all possible configurations of $V(x)$, and $G$ is the matrix inverse of $W$:
\be
\label{Kdef} 
a\sum_{x''} G(x-x'') W(x''-x') = a^{-1}\delta_{xx'} .
\ee

\subsubsection{ITDH approximation}

 We now apply these results to  the ITDH approximation. The particle wave function at  imaginary-time  $\tau$  in a static potential configuration $V (x)$ can be expressed in terms of the energies  $E_n$ and  eigenstates  $\phi_n(x) $  of a particle moving in this potential as 
\be
\label{phiV}
\phi(x, \tau) =  a^{1/2}\frac{R_0(\tau)}{R(\tau)}  \sum_n \phi_n(x) \phi_n^*(x_0) e^{-E_n \tau}  ,
\ee
where the eigenstates are normalized such that
\be
a \sum_x \phi^*_n (x) \phi_{n'}(x) = \delta_{n n'}, \,\,\,  \sum_n \phi^*_n(x)\phi_n(x') =  a^{-1}\delta_{x x'} .
\ee
Then, at large $\tau$, we obtain from (\ref {Calphatau}):
\be
C_{\bm{\alpha}} (\tau) \approx |\phi_B(x_0)|^2 
e^{- E^0_V \tau} ,
\label{largetau}
\ee
where $E^0_V$ is the energy of the lowest energy eigenstate in the potential $V$ that has significant weight at the injection point $x_0$, and $\phi_B$ is the corresponding wave function. Here we have used the fact that in the classical phonon limit, 
$\beta_k = \beta_{0k} = \alpha_k$. Also, since there is an energy gap separating the lowest state in a potential well from all higher energy states, we should be justified in ignoring the contributions of those states at large $\tau$.

Typically, one finds that for the optimum well shape, the minimum excitation energy for a particle in the well will be of the order of $E_V^0$. Thus the condition for (\ref{largetau})  to hold will be $\tau >1/  E_V^0$.  We shall also require that the temperature satisfies $T< E_V^0$.  According to  \cite{HL1}, in the classical phonon limit, one finds $\ln D(E) \sim \tilde{h}( E) / \gamma^2 T$, where $\gamma$ measures the strength of the electron-phonon coupling, and $\tilde{h}<0$ is a function that depends on the energy but is independent of $\gamma$ and $T$. According to (\ref{taufprime}),  the interesting values of  $\tau$ will be related to the energy $E$ by $\tau = \tilde{h}’(E)  / \gamma^2 T$.  Thus $\tau$  will satisfy the inequality $\tau >1/  E_V^0$ if $T$ is sufficiently small, and it will satisfy it for all temperatures of interest ($T<E_V^0$)  if $ \gamma^2 <\tilde{h}’(E )$.

Now, in principle, we should compute $C(\tau)$ from 
 \be
 \label{CtA}
 C(\tau) \to  \int \mathcal{D}V  w_{\rm{cl}} \, |\phi_B (x_0)|^2 e^{-E^0_V \tau} ,
 \ee
 where the integral is over all possible configurations of $V(x)$.
 However, as explained previously, in the low energy tail, the integral will be dominated by configurations close to the one that maximizes the integrand.  Thus, ignoring the correction factor $S(\tau)$ in Eq. (\ref{CCmax}),  
    we may approximate (\ref{CtA})  as
 \be
 \label{Cmax}
 C(\tau) \approx {\rm{max}} \left[ Z_{\rm{cl}} w_{\rm{cl}} \,  |\phi (x_0)|^2 e^{-E_\phi \tau} \right],
 \ee
 where $E_\phi$ is given by (\ref{Ephi}) and the maximum is taken over all choices of $\phi$ as well as of $V(x)$.   If we ignore, for the moment, the factor $|\phi(x_0) |^2 $, then setting $\delta C/ \delta \phi =0$   leads  to the results that $\phi$ is indeed an energy eigenstate in  the potential $V$, which we identify with $\phi_B$ and $E_\phi=E^0_V$.
 Now setting $\delta C /\delta V = 0$ gives the result 
 \be
 \label{Vopt}
 V(x) = V_{\rm{opt}}(x)= - \tau  U(x)  ,
 \ee
  \be
 \label{Udef}
 U(x) \equiv  a \sum_{x'}  W(x-x') |\phi_B (x')|^2  ,
 \ee
 where we made use of Eq. \eqref{Kdef}.
 %These equations have the same form as the ones employed in Ref. \cite{HL1} to determined the optimum shape of the potential and the corresponding trial wave function in their analysis of the density of states tail for a particle in a Gaussian random potential.  
 
 Although Eqs. \eqref{Vopt}-\eqref{Udef} are invariant under translation of the center position of $\phi$ and do not specify it, maximizing the prefactor in \eqref{Cmax} dictates that we choose $\phi(x)$ to be centered at $x_0$.
 However, other than this, taking into account contributions to the variational derivative from the factor  $|\phi(x_0) |^2 $ would change these results by an amount that is small  and can be neglected in the low-energy tail.  
 
For an arbitrary wave function $\phi$ centered at $x
_0$ and an arbitrary potential $V$ we may define a smoothed potential by
\be
\label{Vsmooth}
V_s(x_0) = a \sum_{x}  \abs{\phi (x)}^2 V(x) ,
\ee
Let us define $V_s^{ {\rm{opt}} }$ as the value of 
$V_s(x_0)$ when $V = V_{\rm{opt}}$ and $\phi = \phi_B $.
Let  $\theta = \squish {\phi_B} {H_e} {\phi_B}$ be the particle kinetic energy in the state $\phi_B $.
Then at the optimum point, we have 
\be
E_\phi = \theta +  V_s^{ {\rm{opt}} }.
\ee
Using \eqref{wcl}, \eqref{Cmax} and (\ref{Vopt}), we then obtain
\be
\label{CclITDH}
C(\tau) \approx |\phi_B (x_0)|^2 e^{-  \tau (\theta + V_s^{ {\rm{opt}} }) }     e^{ - \tau^2  \sigma_0^2 /2} ,
\ee
where 
\be
\label{sig0}
\sigma_0^2 =  a^2 \sum_{x x'} \abs{\phi_B (x)}^2 W(x-x') \abs{\phi_B (x')}^2 .
\ee
Also, we have $V_s^ {\rm{opt}}= - \tau \sigma_0^2$.

\subsubsection{FWF approximatiion}
 
We now turn to the FWF approximation.  In the classical phonon limit,  the exponent $F(\tau)$ in  Eq. (\ref{Fdef})  may be written as 
\be
F(\tau) = \tau^2 \sigma_{\phi_{\rm{tr}}}^2 / 2
\ee
with  
\be
\sigma_{\phi_{\rm{tr}}}^2 = 2T\sum_k \frac{|C_k |^2} {\omega_k}
=
a^2 \sum_{x x'} \abs{\phi_{\rm{tr}}  (x)}^2 W(x-x') \abs{\phi_{\rm{tr}}  (x')}^2,
\label{sigmatr}
\ee
where we made use of Eq. \eqref{Wclass}.
% The right-hand side of this equation remains finite in the classical phonon limit, and it is equal to the quantity $\sigma _0^2$ defined by (\ref{sig0}), with $\phi$ replaced by $\phi_{\rm{tr}}$.
Also, we may identify $E_e^\phi$ with $\theta$.  Then, Eq. (\ref{CFC}) becomes
\be
\label{Cprt}
C_{\phi_{\rm{tr}}} (\tau) \approx  |\phi_{\rm{tr}} (x_0)|^2 e^{- \tau \theta} e^{\tau^2  \sigma_{\phi_{\rm{tr}}}^2 /2 }.
\ee
In order to find the trial wave function which maximizes this function, we ignore the pre-exponential factor and set equal to zero the variational derivative of the exponent with respect to $\phi_{\rm{tr}}(x)$.  This gives the equation
\be 
H_e \phi_{\rm{tr}}(x) + V_\phi (x) \phi_{\rm{tr}}(x) = \mu \phi_{\rm{tr}}(x) ,
\ee
\be
V_\phi (x) \equiv - \tau a  \sum_{x'} W(x-x') |\phi_{\rm{tr}}(x)|^2 ,
\ee
where $\mu$ is a Lagrange multiplier necessary to enforce the constraint that $\phi_{\rm{tr}}$
is properly normalized.  We see that these are the same equations as the ones we used to determine the optimum potential $V_{\rm{opt}}$ and the corresponding wave function $\phi$ in the ITDH. Moreover, when we identify $\phi_{\rm{tr}}$ with the wave function $\phi$ obtained in the ITDH approach, we find that 
$\sigma_{\phi_{\rm{tr}}} = \sigma_0$ and 
the value of $C(\tau)$ obtained from (\ref{Cprt}) coincides with (\ref{CclITDH}).

Thus, the FWF and ITDH approximations give equivalent predictions in the regime under discussion. We shall see that these results also agree, as far as the exponential factors are concerned, with the predictions of \cite{HL1} for the low-energy tail of the density of states in a Gaussian random potential.

\section{Continuum with acoustic phonons}
\label{continuum} 

For a system of particles interacting with acoustic phonons, a continuum model  can be considered
when the relevant particle wavelength is far larger than the lattice spacing $a$. The Hamiltonian for such a system is 
\begin{align}
    H=\int dx &\left(\psi^\dag(x)\left(-\frac{1}{2m}\frac{d^2}{dx^2}\right)\psi(x)+\gamma \frac{\partial u}{\partial x} \psi^\dag(x)\psi(x)\right.\notag
    \\
    &\left.\quad+\frac{1}{2}K\left(\frac{\partial u}{\partial x}\right)^2+\frac{1}{2\rho}\Pi(x)^2\right),
    \label{continuumHam}
\end{align}
where $\psi(x)$ and $\psi^\dag(x)$ are particle annihilation and creation operators at a position $x$, with $\comm{\psi(x)}{\psi^\dagger (x')}_{\mp}= \delta (x-x')$ ($-$ for bosons, $+$ for fermions),  $m$ is the particle mass, $\gamma$ is the particle-phonon coupling strength, $u(x)$ is the nuclear displacement field, $K$ is a bulk modulus, $\rho$ is nuclear mass density, and $\Pi$ is nuclear momentum density (momentum per unit length).
The strain $\varepsilon(x)\equiv\frac{\partial u(x)}{\partial x}$ is assumed to be small, $\abs{\varepsilon(x)}\ll1$, and the commutation relation for the nuclear displacement and momentum density is
\begin{align*}
    \comm{u(x)}{\Pi(x')}=i\delta(x-x').
\end{align*}
The phonon dispersion relation for the one-dimensional harmonic chain is given in the  continuum limit  by 
\begin{equation}
    \omega_k=\frac{2}{a} \left(\frac{K}{\rho}\right)^{1/2}   \sin\frac{|k|a}{2} \, \to \, |k| \left(\frac{K}{\rho}\right)^{1/2} .
    \label{dispersion}
\end{equation}
Note the Hamiltonian in Eq. \eqref{continuumHam} is the special case of the more general Hamiltonian in Eq. \eqref{generalH} with  $\omega_k$ given by  Eq. \eqref{dispersion}, $\epsilon(k)=\frac{k^2}{2m}$, and $\gamma_k=\gamma\abs{k}$.

\subsection{Classical self-trapping ground state}
In the classical phonon limit $\rho\to\infty$ while keeping the other parameters fixed, the Hamiltonian \eqref{continuumHam} becomes
\begin{align}
    H=\int dx &\left(\psi^\dag(x)\left(-\frac{1}{2m}\frac{d^2\psi(x)}{dx^2}\right)+\gamma \varepsilon(x)\psi^\dag(x)\psi(x)\right.\notag
    \\
    &\quad\left.+\frac{K}{2}\varepsilon(x)^2\right),
    \label{continuumHamCl}
\end{align}
where $\varepsilon (x)$ may be considered as a fixed classical function of position.  

We wish to find a strain configuration $\varepsilon(x)$ and a single-particle wave function $\phi(x)$ which gives the lowest possible  expectation value of the Hamiltonian in the single-particle subspace.  
First, the minimum with respect to $\varepsilon(x)$ is found by taking the functional derivative
\begin{align*}
    \frac{ \delta \sex{ H}}{\delta \varepsilon(x)}
    =
    \gamma |\phi(x)|^2 +K\varepsilon(x)
    =
    0.
\end{align*}
Thus, we require
\begin{align}
    \varepsilon(x)=-\gamma \abs{\phi(x)}^2/K.
    \label{LatRelaxSt}
\end{align}
Note that larger local particle density $\abs{\phi(x)}^2$ induces more  strain $\varepsilon(x)$.  
 Substituting this back into the Hamiltonian \eqref{continuumHamCl} in the single particle subspace, one obtains a lattice-relaxed total energy
\begin{align}
    \ev{H}_{\rm{LR}}
    =
    \int dx \left(\frac{1}{2m}\abs{\frac{d\phi(x)}{dx}}^2
    -\frac{\gamma^2\abs{\phi}^4}{2K}\right),
    \label{Hrelaxed}
\end{align}
where integration by parts was used for the kinetic energy term.
Next, we minimize $\ev{H}_{\rm{LR}}$ with respect to $\phi^*(x)$ with the constraint $\int dx\abs{\phi}^2=1$ using Lagrange multiplier $\mu$:
\begin{align}
    &\frac{\delta}{\delta \phi^*(x)}\left(\ev{H}_{\rm{LR}}-\mu\int dx\abs{\phi(x)}^2\right)\notag
    \\
    &\quad=
    -\frac{1}{2m}\frac{d^2\phi(x)}{dx^2}
    -\frac{\gamma^2\abs{\phi(x)}^2\phi(x)}{K}
    -\mu\phi(x)
    =
    0.\label{Hminpsi}
\end{align}
This is a nonlinear Schr\"odinger equation for a particle in the effective potential $-\frac{\gamma^2\abs{\phi}^2}{K}$ with particle energy eigenvalue $\mu$. 
Since any position-dependent phase factor of $\phi(x)$ leads to higher kinetic energy, the ground state should have a global (position-independent) phase factor. Thus, the ground state can be chosen to be real and Eq. \eqref{Hminpsi} becomes
\begin{align}
    -\frac{1}{2}\frac{d^2\phi(x)}{dx^2}
    -\frac{1}{2}\nu\phi(x)^3
    =
    -\frac{\kappa_H^2}{2}\phi(x),
    \label{HLnonlinSE}
\end{align}
where $\nu=\frac{2m\gamma^2}{K}$ and $\kappa_H=\left(-2m\mu\right)^{1/2}$.
Since Eq. \eqref{HLnonlinSE} is translationally invariant, it has degenerate solutions, which have the form (cf. Ref. \cite{HL1} Appendix B) 
\begin{align}
    \phi(x)=(\kappa_H/2)^{1/2}\sech(\kappa_H (x-x_0))
    \label{HLpsi}
\end{align}
with arbitrary center  position $x_0$. The wavefunction in Eq. \eqref{HLpsi} is normalized and its normalization condition gives
\begin{align}
    \kappa_H=\nu/4=\frac{m\gamma^2}{2K}.
\end{align}
The particle energy of the ground state is
\begin{align}
    \mu=-\frac{\kappa_H^2}{2m}=-\frac{m\gamma^4}{8K^2},
\end{align}
and the total energy of the ground state (minimum energy) is obtained by substituting \eqref{HLpsi} into \eqref{Hrelaxed}:
\begin{align}
    E_{\rm{min}}=-\frac{\kappa_H^2}{6m}=-\frac{m\gamma^4}{24K^2}.
    \label{EminCl}
\end{align}
Note that the difference between the total and particle energies,
\begin{align}
    E_{\rm{min}}-\mu=\frac{\kappa_H^2}{3m}=-\frac{m\gamma^4}{12K^2},
\end{align}
is the elastic energy.

We remark that if $\rho$ is not infinite, the quantum ground state of the system with one particle present will not be localized but will be a polaron with total momentum $k=0$. This state may be considered as a linear superposition of the self-trapped particle with arbitrary center positions $x_0$.
In the limit $\rho\to\infty$, polaron energy approaches $E_{\rm{min}}$ \eqref{EminCl} and the energy becomes independent of the polaron momentum as the polaron mass becomes infinite.

In summary, in the classical phonon limit $\rho\to\infty$, the ground state for the system of a single particle interacting with acoustic phonons in one dimension  is described by a self-trapped particle state \eqref{HLpsi} with lattice relaxation \eqref{LatRelaxSt}. The self-trapping energy $|E_{\rm{min}}|$ is non-zero for any 
non-zero $\gamma$, although it can be very small when $\gamma$ is small.

\subsection{Application of the fixed-wave-function approximation}

Now the FWF approximation introduced in Sec. \ref{FWF} can be implemented.
We choose a trial wave function of the form 
\begin{align}
    \phi_\kappa(x)=(\kappa/2)^{1/2}\sech(\kappa x),
    \label{phitr}
\end{align}
with a single variational parameter $\kappa$. 
This seems like a good choice, since it is  the correct  form  in the classical phonon limit $\rho \to \infty$, where it  coincides with the results of \cite{HL1} for the Gaussian white noise potential, and it happens to coincide with the form of the self-trapped ground state at $T=0$. 
We will eventually choose $\kappa$ to maximize
 the correlation function $C$ as in Eq. \eqref{Cmaximization} [or to maximize the density of states $D(E)$ as in Eq. \eqref{maxDOS} in the modified FWF approximation, c.f., Sec. \ref{modifiedaprx}].
In the  continuum, Eq. \eqref{Fdef} becomes
\begin{align}
    F(\tau)
    =
    \int_{-\pi/a}^{\pi/a} \frac{dk}{2\pi} 
    \frac{L \abs{C_k}^2}{\omega_k^2} [(n_k&+1)e^{-\omega_k\tau}+n_k e^{\omega_k\tau}\notag
    \\
    &+\omega_k\tau-(2n_k+1)],
    \label{Fdefint}
\end{align}
where $C_k$ is defined in Eq. \eqref{eq:Ck}, with
\begin{align*}
    \lambda_k&=\sqrt{\frac{1}{2\rho L \omega_k}}\gamma\abs{k},
    \\
    \abs{\phi^2}_k
    &=
    \int dx e^{-ikx} \abs{\phi_\kappa(x)}^2
    =
    \frac{k\pi}{2\kappa}\csch(\frac{k\pi}{2\kappa}).
\end{align*}
Note that $\abs{\phi^2}_k$ is a smooth function, with $\abs{\phi^2}_k\to1$ for $k\ll \kappa$ and $\abs{\phi^2}_k\to0$ for $k\gg\kappa$, and $\abs{\phi^2}_k$ is significant only for $\abs{k}\lesssim\kappa$.
Furthermore,
\begin{align}
    \phi_\kappa(x_0)&=(\kappa/2)^{1/2}\text{ for }x_0=0,
    \\
    E_e^{\phi_\kappa}
    &=\mel{\phi_\kappa}{H_e}{\phi_\kappa}
    =
    \frac{\kappa^2}{6m}
    \label{sechKin}.
\end{align}
Thus, using the FWF approximation, $C_{\phi_\kappa}(\tau)$ [Eq. \eqref{CFC}] is essentially obtained by performing the numerical integration in  Eq. \eqref{Fdefint}. Then, $C_{\phi_\kappa}(\tau)$ is maximized with respect to $\kappa$ [Eq. \eqref{Cmaximization}] and the density of states $D(E)$ is obtained from $C(\tau)$ by inverting the Laplace transform \eqref{laplace} as described in Sec. \ref{invert}.

We first consider two special limits, the  classical phonon limit and the quantum zero temperature, 
in order to obtain further insights.

\subsubsection{Classical phonon limit}
The classical phonon limit is achieved by taking $\rho\to\infty$ while keeping the temperature $T$ fixed. This reduces to the case of  a particle  in a static Gaussian white noise potential, which was considered in Halperin--Lax analysis \cite{HL1,HL2} where the optimal $\kappa$ maximizing the density of states as in Eq. \eqref{maxDOS} was given by $\kappa_E=(-2mE)^{1/2}$. We would like to compare this Halperin--Lax result to the FWF approximation.

In the classical phonon limit, $\omega_k\to0$ [c.f. Eq. \eqref{dispersion}] and $\omega_k\tau\ll1$ which simplifies Eq. \eqref{Fdefint} to
\begin{align*}
    F(\tau)
    =
    \int_{-\pi/a}^{\pi/a} \frac{dk}{2\pi/L} 
    \abs{C_k}^2
    (n_k+1/2)\tau^2.
\end{align*}
Since $\omega_k/T\ll1$, the above expression is further reduced to
\begin{align}
    F(\tau)
    \approx
    \frac{\gamma^2 T}{2\rho}
    \int_{-\pi/a}^{\pi/a} \frac{dk}{2\pi} 
    \frac{k^2}{\omega_k^2}
    \abs{\phi^2}_k^2
    \tau^2.
    \label{FtauClAprx}
\end{align}
The integrand of Eq. \eqref{FtauClAprx} is significant for $\abs{k}\lesssim\kappa$, due to $\abs{\phi^2}_k$, and in this interval, $\omega_k$ can be approximated as a linear dispersion $\omega_k\approx\sqrt{K/\rho}\abs{k}$, since $\kappa a\ll1$ in the continuum limit.
In addition, for $\kappa a\ll1$, the integral can be evaluated, since
\begin{equation}
    \int_{-\pi/a}^{\pi/a} \frac{dk}{2\pi} 
    \abs{\phi^2}_k^2
    \approx
    \int_{-\infty}^{\infty} \frac{dk}{2\pi} 
    \abs{\phi^2}_k^2
    =
    \kappa/3.
\end{equation}
Thus, one obtains
\begin{align}
    F(\tau)
    \approx
    \frac{\gamma^2 T}{2K}
    \frac{\kappa}{3}
    \tau^2
    \equiv \sigma_{\phi_\kappa}^2\tau^2/2
    \label{classicalFtau}.
\end{align}
Thus, from Eq. \eqref{CFC}, one gets
\begin{align}
    C_{\phi_\kappa}(\tau)= \abs{\phi_\kappa(x_0)}^2e^{-\tau E_e^{\phi_\kappa}+\sigma_{\phi_\kappa}^2\tau^2/2}.
\end{align}

Now the correlation function must be maximized with respect to  the variational parameter $\kappa$.  The value of
$\kappa$ that maximizes the correlation function is found by solving the eqution 
\begin{align}
    \frac{d\ln C_{\phi_\kappa}(\tau)}{d\kappa}
    =
    \frac{d}{d\kappa}
    \left[
    \ln\kappa
    -\tau E_e^{\phi_\kappa}+\sigma_{\phi_\kappa}^2\tau^2/2\right]
    =0,
    \label{kappamaxeq}
\end{align}
which, for $\kappa>0$, gives
\begin{align}
    \kappa_\tau=\frac{m\gamma^2 T\tau}{4K}+\sqrt{\left(\frac{m\gamma^2 T\tau}{4K}\right)^2+\frac{3m}{\tau}}.
    \label{kappaopt}
\end{align}

The asymptotic form of $D(E)$ can now be obtained using the inversion formulas described in Sec. \ref{invert}. 
Using the optimal $\kappa$, given by (\ref{kappaopt}),  Eq. \eqref{EdlnCdt} gives
\begin{align}
    E
    =
    -\frac{\kappa^2}{6m}-\sqrt{\left(\frac{\kappa^2}{3m}\right)^2-\frac{2\gamma^2 T}{3K}\kappa},
    \label{Ekapparel}
\end{align}
For low energy $E\ll0$, one obtains  
the Halperin--Lax result $\kappa=(-2mE)^{1/2}$ [c.f. Eq. \eqref{Ekapparel}].
 The resulting approximation for $D(E)$ is 
\begin{align}
    D_{\rm{FWF}}(E)
    \approx
    \left(\frac{\kappa_E}{2\pi \xi}\right)^{1/2}e^{-\frac{4\kappa_E^3}{3m^2\xi}}&\text{ for }E\ll0,
    \label{DOSclassical}
\end{align}
where $\xi$ is introduced to make connection to the Gaussian white noise model [c.f. Eq. \eqref{GWNcontinuumParams}]
\begin{align}
    \frac{\xi}{2}=\frac{\gamma^2T}{K}.
\end{align}
[See, also,  the discussion in Appendix \ref{DOSmax} of the modified FWF approximation, which maximizes $D(E)$ rather than the correlation function $C(\tau)$.]

\begin{figure}
    \centering
    \includegraphics[width=3in]{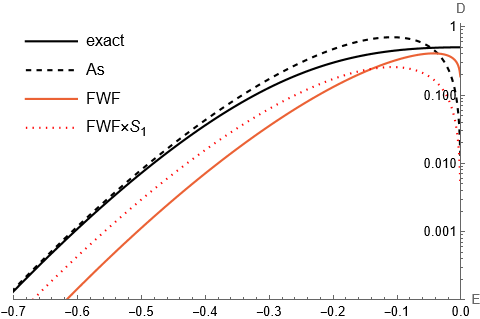}
    \caption{Comparison of different approximations to the densitiy of states in the Gaussian white noise potential. The exact $D_{\rm{exact}}(E)$ \eqref{rhoGWNexact} and exact  asymptotic form \eqref{DOSGWNAS} are denoted as exact (black solid curve) and As (black dashed curve), respectively. The FWF asymptotic form \eqref{DOSclassical} and its form with the higher-order correction \eqref{DOSclassicalS1} are denoted as FWF (red solid curve) and FWF$\times S_1$ (red dotted curve), respectively. We use $K=1$, $m=1$, $T=0.1$, and $\gamma=1$.}
    \label{fig:GWNcomp}
\end{figure}

Our approximate  density of states from classical acoustic phonons $D_{\rm{FWF}}(E)$ can be compared to the exact density of states for the Gaussian white noise potential \cite{Halperin65,HL1}
\begin{align}
    D_{\rm{exact}}(E)&=(m^2\xi)^{-1/3}N'(E(m^2\xi)^{-2/3}),
    \label{rhoGWNexact}
\end{align}
where $N$, which may be expressed in terms of Airy functions as
\begin{align}
    N(\epsilon)&=\pi^{-2}([\mathrm{Ai}(-2\epsilon)]^2+[\mathrm{Bi}(-2\epsilon)]^2)^{-1},
\end{align}
is the cumulative density of states as a function of unitless energy $\epsilon=Em^{-4/3}\xi^{-2/3}$. This gives rise to the exact asymptotic form,
\begin{align}
    D_{\rm{As}}(E)
    &\approx
    \frac{4\kappa_E^2}{\pi m\xi}e^{-\frac{4\kappa_E^3}{3m^2\xi}}\text{ for }E\ll0.
    \label{DOSGWNAS}
\end{align}
\arraybackslash
Note Eqs. \eqref{DOSclassical} and \eqref{DOSGWNAS} have the same exponential factor, but different prefactors.
The difference in the prefactors can be reduced by considering higher-order corrections, as discussed in Appendix \ref{prefactors}.
The correction factor $S_1(E)$ is given in Eq. \eqref{S1factor}, which gives corrected density of states to FWF approximation:
\begin{align}
    D_{\rm{FWF}}(E) \, S_1(E)
    \approx
    \left(\frac{2}{15}\right)^{1/2}
    \frac{4\kappa_E^2}{\pi m\xi}
    e^{-\frac{4\kappa_E^3}{3m^2\xi}}&\text{ for }E\ll0.
    \label{DOSclassicalS1}
\end{align}
The comparison of different approximations to the density of states in the Gaussian white noise potential is given in Fig. \ref{fig:GWNcomp} where we use $K=1$, $m=1$, $T=0.1$, and $\gamma=1$. 

\subsubsection{Quantum zero temperature}
Zero temperature can be considered instead of the classical phonon limit. At zero temperature $T=0$, all Bose occupation numbers $n_k$ are set to zero.  Then, Eq. \eqref{Fdefint} becomes
\begin{align*}
    F_{T=0}(\tau)
    =
    \int_{-\pi/a}^{\pi/a} \frac{dk}{2\pi} 
    \frac{L \abs{C_k}^2}{\omega_k^2} [e^{-\omega_k\tau}+\omega_k\tau-1].
\end{align*}
For  $\omega_k\tau\ll1$, $F_{T=0}(\tau)$ is quadratic in $\tau$ 
\begin{align}
    F_{T=0}(\tau)
    &\approx
    \int_{-\pi/a}^{\pi/a} \frac{dk}{4\pi} 
    L \abs{C_k}^2
    \tau^2
    \\
    &\approx
    \frac{3\gamma^2\zeta(3)\kappa^2}{2\pi^3 (\rho K)^{1/2} } \,
    \tau^2,
    \label{FT0smalltau}
\end{align}
where similar approximations were used as in Eq. \eqref{classicalFtau}, and $\zeta$ is the Riemann zeta function.
% $\kappa$ that maximizes the correlation function is found by
% \begin{align}
%     \frac{d\ln C_{\phi_\kappa}(\tau)}{d\kappa}
%     =
%     \frac{d}{d\kappa}
%     \left[
%     \ln\kappa
%     -\tau E_e^{\phi_\kappa}+Q_\kappa^2\tau^2/2\right]
%     =0,
%     \label{kappamaxeq2}
% \end{align}
For $\omega_k\tau\gg1$,
\begin{align}
    \frac {d F_{T=0}(\tau)}{d \tau}
    &\approx
    \int_{-\pi/a}^{\pi/a} \frac{dk}{2\pi} 
    \frac{ L\abs{C_k}^2}{\omega_k}
    \approx
    \frac{\gamma^2\kappa}{6K},
\end{align}
where, again,  similar approximations were used as in Eq. \eqref{classicalFtau}.
According to  Eq. \eqref{EdlnCdt}, this implies that for a fixed value of $\kappa$, the computed density of states
vanishes below a  minimum energy 
\begin{align}
    E_{\rm{min}, \kappa}=E_e^{\phi_\kappa}-\frac{d F_{T=0}(\tau)}{d\tau}
    =
    \frac{\kappa^2}{6m}-\frac{\gamma^2\kappa}{6K},
    \label{EminEq}
\end{align}
which implies that $D(E)=0$ for 
\be
E < E_{\rm{min}}=  {\rm{min}}
_\kappa (E_{\rm{min}, \kappa}) = - \frac {m \gamma^4}{24 K ^2} .
\ee
This agrees with the result (\ref {EminCl}) for the ground-state energy of the self-trapped particle in the limit $\rho \to \infty$. 

We note that in  Eq. 
\eqref{EminEq}, the first term is the kinetic energy of the FWF
[c.f. Eq. \eqref{sechKin}], and the second term is the 
particle-phonon interaction and the elastic energy of the FWF after lattice relaxation [c.f. the second term of Eq. \eqref{Hrelaxed} for $\phi=\phi_\kappa$].

\subsection{Numerical results}
We now turn to situations where neither $T$ nor $\rho^{-1}$ is zero. Here we use numerical methods to compute $F(\tau)$ for various choices of the parameter $\kappa$ and to find the value of $\kappa$ that maximizes the resulting estimate of $C(\tau)$.  Finally, we use the method of Sec. \ref{invert} to obtain the density of states $D(E)$. 

The ratio of the thermal energy to the energy of a phonon
% relevant to the localization 
$\alpha=T/\omega_{\kappa_H}$ is a useful parameter for understanding the importance of  quantum phonon effects. The limit $\alpha\gg1$ will approach  classical phonon limit, and small $\alpha$ will show the effects of quantum motion of the nuclei. 

To study how the density of states depends on this parameter $\alpha$, we can either 1) vary $\rho$ while fixing the temperature $T$, or 2) vary $T$ while fixing $\rho$. We use $a=0.1$, $K=1$, and $m=1$, and we vary $\rho$, $T$, and $\gamma$ for the following calculations.  
Equivalently, we are measuring $E$ in units of $\left(\frac{K^2}{m}\right)^{1/3}$, and $D(E)$  in units of $\left(\frac{m^2}{K}\right)^{1/3}$.

\subsubsection{\texorpdfstring{Varying $\rho$ while fixing $T$}{Lg}}
\begin{figure}
    \centering
    \includegraphics[width=3in]{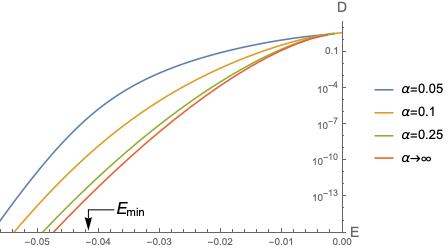}
    \caption{Densities of states in the FWF approximation, $D_{\rm{FWF}}(E)$,  with trial wave functions of the form (\ref{phitr}), for different $\alpha$ values achieved by varying $\rho$ for fixed $T=\SI{5e-4}{}$, and $\gamma=1$. Parameters $m$ and $K$ have been set equal to unity. 
    $E_{\rm{min}}$ is the minimum energy \eqref{EminCl} of the self-trapped particles at $T=0$, for $\rho \to \infty$.}
    \label{fig:gamma1Tfixed}
\end{figure}

\begin{figure}
    \centering
    \includegraphics[width=3in]{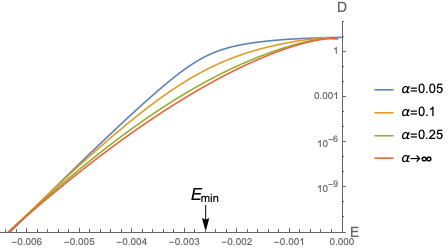}
    \caption{Densities of states $D_{\rm{FWF}}(E)$ for different $\alpha$ values achieved by varying $\rho$ for fixed $T=\SI{1.25e-4}{}$, and $\gamma=0.5$. Other parameters are the same as in Fig. \ref{fig:gamma1Tfixed}.}
    \label{fig:gamma05Tfixed}
\end{figure}

The dependence of the density of states on $\alpha$ by changing $\rho$ is shown in Figs. \ref{fig:gamma1Tfixed} and \ref{fig:gamma05Tfixed} for $\gamma=1$ and $0.5$, respectively. For smaller $\alpha$, the density of states is larger, meaning the nuclear quantum effect increases the density of states.
The limit $\alpha\to\infty$, achieved by $\rho\to\infty$, corresponds to the classical phonon limit considered by Halperin and Lax \cite{HL1}.
It is also seen that the different $\gamma$ values change the overall energy scale.


\subsubsection{\texorpdfstring{Varying $T$ while fixing $\rho$}{Lg}}
\begin{figure}
    \centering
    \includegraphics[width=3in]{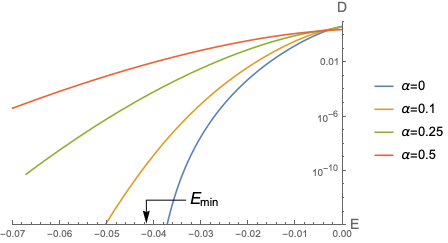}
    \caption{Densities of states $D_{\rm{FWF}}(E)$ for different $\alpha$ values achieved by varying $T$, for fixed $\rho=\SI{e4}{}$, and $\gamma=1$. Other parameters are the same as in Fig. \ref{fig:gamma1Tfixed}. }
    \label{fig:gamma1rho4}
\end{figure}

\begin{figure}
    \centering
    \includegraphics[width=3in]{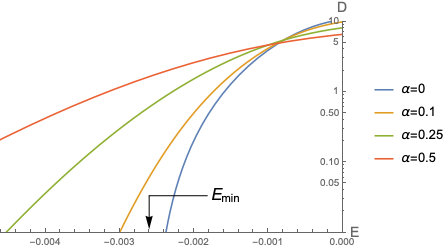}
    \caption{Densities of states $D_{\rm{FWF}}(E)$  for different $\alpha$ values achieved by varying $T$, for fixed $\rho=\SI{e4}{}$, and $\gamma=0.5$. Other parameters are the same as in Fig. \ref{fig:gamma1Tfixed}.}
    \label{fig:gamma05rho4}
\end{figure}

The dependence of the density of states on $\alpha$ by changing $T$ is shown in Figs. \ref{fig:gamma1rho4} and \ref{fig:gamma05rho4} for $\gamma=1$ and $0.5$, respectively. For larger values of $\alpha$, the density of states increases, which implies that a higher temperature leads to a larger density of states.
Note at zero temperature $\alpha=0$, there exists energy minimum given by Eq. \eqref{EminEq} below which the density of states completely vanish.
Comparing with Figs. \ref{fig:gamma1Tfixed} and \ref{fig:gamma05Tfixed}, we see that the curves for $\alpha=0.25$ and $\alpha=0.5$ would be almost indistinguishable from the classical phonon limit $\rho\to\infty$ at the given temperatures, which was discussed in \cite{HL1}.

If we restore the parameters $m$ and $K$, the density of states can be written in the form
\be
D(E,T,\rho,\gamma) = \left( \frac {m}{E_\gamma} \right)^{1/2} \tilde{D} , 
\ee
where $E_\gamma \equiv \gamma^4 m K^{-2} = 24 |E_{\rm{min}}|$, and $\tilde D$ depends only on the dimensionless variables $E/E_\gamma ,  \,  T / E_\gamma $ and $\rho E_\gamma / mK$. This means that plots of $D(E,T,\rho,\gamma) $ for $\gamma =0.5$ and $\gamma =1$ would coincide if we change variables accordingly. I.e.,
\begin{align}
    D(E,T,\rho,\gamma=0.5)=4D(E',T',\rho',\gamma=1)
\end{align}
with $E'=16E,T'=16T,\rho'=\rho/16$. The value of $D(E,T,\rho,\gamma)$ is independent of the remaining dimensionless quantity $\tilde {\gamma} \equiv \gamma^3 m K^{-2}$ because if we rescale the field $u(x)$ by a factor of $\lambda$, the coefficients in the Hamiltonian (\ref{continuumHam}) will be modified and $\tilde\gamma$ will be changed by a factor of $\lambda$, but the energy eigenvalues are unchanged.
However, $du/dx$ would no longer be the strain. The Hamiltonian \eqref{continuumHam} will actually become unphysical for sufficiently large values of $\gamma$, because the resulting strains can be larger than 1.

\subsubsection{\texorpdfstring{Relation between $\kappa$ and energy}{Lg}}
\begin{figure}
    \centering
    \includegraphics[width=3in]{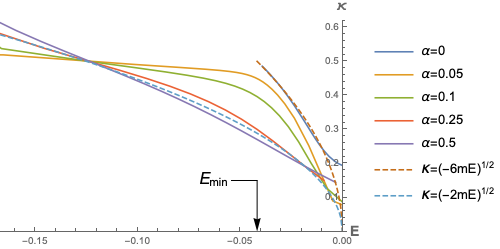}
    \caption{Energy $E$ and optimal $\kappa$ relation for different $\alpha$ values achieved by varying $T$ for fixed $\rho=\SI{e4}{}$, and $\gamma=1$.
    $\alpha=0$ (zero temperature) curve follows $\kappa=(-6mE)^{1/2}$, and
    $\alpha=0.25$ and $0.5$ curves follow classical phonon limit result,  $\kappa=(-2mE)^{1/2}$.}
    \label{fig:Ekappa}
\end{figure}

Fig. \ref{fig:Ekappa} shows the relation between energy $E$ and optimal $\kappa$ that maximizes correlation function $C_{\phi_{\rm{tr}}}(\tau)$  for different $\alpha$ values, achieved by varying $T$ for fixed $\rho=\SI{e4}{}$, and $\gamma=1$.
We find that the $\alpha=0$ (zero temperature) curve follows $\kappa=(-6mE)^{1/2}$, for energies above the minimum energy $E_{\rm{min}}$, while the curves for  $\alpha=0.25$ and $0.5$ are close to the  classical phonon limit result,  $\kappa=(-2mE)^{1/2}$. The curves for $\alpha = 0.05$ and $\alpha = 0.1$ exhibit a more complicated behavior. 

We can understand the result  $\kappa = (-6mE)^{1/2}$ for $E_{\rm{min}}<E<0$ for $\alpha=0$, and $\kappa = (-2mE)^{1/2}$ for classical phonon limit (large $\alpha$) as follows. 
Ignoring $\abs{\phi_\kappa(x_0)}^2$ factor in Eq. \eqref{CFC}, one obtains
\begin{align}
    C_{\phi_\kappa}(\tau)\approx
    e^{-\tau E_e^{\phi_\kappa}+F(\tau)},
\end{align}
where, according to Eq. \eqref{FT0smalltau}, at $T=0$, for $\tau$ in the range of interest, $F$ has the form 
\begin{align}
    F(\tau)\approx A\kappa^p \tau^2.
\end{align}
with $A$ a constant and $p=2$. 
In the classical phonon limit, $F$ has a similar form but with $p=1$. [C.f. Eq. \eqref{classicalFtau}]. 
In either case, the optimal $\kappa$ that maximizes $C$ can be obtained from
\begin{align}
    \frac{d\ln C_{\phi_\kappa}(\tau)}{d\kappa}
    \approx
    -\tau \frac{\kappa}{3m} +p A\kappa^{p-1}\tau^2
    =0.
    \label{kappamaxeq3}
\end{align}
For this optimal $\kappa$,  Eq. \eqref{EdlnCdt} gives the relation between $E$ and $\kappa$:
\begin{align}
    E
    =
    \frac{\kappa^2}{6m}-2A\kappa^p\tau
    =
    \frac{\kappa^2}{m}\left[\frac{1}{6}-\frac{2}{3p}\right].
    \label{Ekapparel3}
\end{align}
This leads to the result 
 $\kappa=(-6mE)^{1/2}$
for $T=0$ ($p=2$)
as well as the known result 
$\kappa=(-2mE)^{1/2}$ for the classical phonon limit ($\alpha$ large, $p=1$).

\section {Additional remarks}
\label{additional}

\subsection {Momentum-dependent spectral density}
\label{AkE} 

Using either the ITDH or FWF approximation, one finds that for a given energy $E$ in the low-energy tail, the tunneling density of states is dominated by a particle wave function of the form 
$\phi(x) = f (x- x_0)$, where $f$ has a fixed shape, and  $x_0$  the position of a local minimum of the fluctuating potential. 
As noted in Refs. \cite{HL1,HL2},  this suggests a simple approximation for the momentum-dependent spectral density
\be
\label{Aksmall}
A^{(1)} (k,E) \approx | \tilde {f} (k)|^2  D(E) ,
\ee 
where $\tilde{f}$ is the Fourier transform of $f(x)$:
\be
\tilde{f} (k) = \int dx e^{-ikx} f(x) .
\ee
(Here we use the continuum normalization, with $\int |f|^2 dx = 1$). 
As in the case of a random potential due to impurities, however, this approximation breaks down if $k$ becomes too large.   Because $f(x)$ is an analytic function of position, its Fourier transform will fall off exponentially for $|k | > l^{-1}$, where $l$ is a measure of the spatial width of the wave function.  Then for sufficiently large $k$, we can obtain a larger contribution to the spectral density from processes where the injected particle emits or absorbs a phonon with  wave vector $\approx \pm k$ in order to bring it into the momentum region where $A^{(1)}$ is largest. The contributions of these processes to the imaginary part of the particle self-energy may be written as 
\begin{eqnarray}
\label{Akbigq} 
\Im \Sigma (k,E) & = & \frac{L}{2\pi^2} \int dk'
|\lambda_{k'}|^2  \nonumber \\  & &  [(n_{k'}+1 ) A^{(1)}(k-k', E-\omega_{k'} ) \nonumber \\
&+& n_{-k'} A^{(1)}(k-k', E+\omega_{-k'}) ] .
\end{eqnarray} 
Because the integrand will be significant only when $|k'-k| \leq \kappa \ll |k|$, we may replace $k'$ by $k$ in $\omega_{k'}$, etc., and we may bring these factors outside the integral. 
This leads to a contribution 
\begin{align}
    \label{Akbig}
    A^{(2)}(k,E)  &= \pi^{-1} \Im [ E -\epsilon_k - \Sigma (k,E) ]^{-1}  \nonumber \\
    &= 
    \frac {\gamma_k ^2 [ (1 + n_k)   D(E - \omega_k) + n_k D(E+ \omega_k)] } {2 \rho \omega_k (\epsilon_k - E)^2    }  .
\end{align}
For intermediate values of $k$, we should approximate the spectral density by the larger of (\ref{Aksmall}) and (\ref{Akbig}). 

Approximation (\ref{Akbig}) will be particularly important in the case of an indirect absorption edge.   For a semiconductor with an indirect band gap, such as Si, where the exciton binding energy is small and the electron-phonon interaction is too weak to produce self-trapping, the low-energy tail of the indirect optical absorption edge can be reproduced by an analogous formula involving a transition of an electron from a  state near the valence band maximum, to  a state near the conduction band minimum,  with emission or absorption of a single phonon \cite{Giustino17}.

\subsection {Three-dimensional systems}
\label{3D}

Three-dimensional systems can differ from one-dimensional systems in several ways, which may require adjustments to the methods described above. One important difference is that in three dimensions, it is necessary for the particle-phonon coupling  to exceed a critical value in order for self-trapping to occur in limit of large nuclear masses, whereas in one-dimension, self-trapping occurs for arbitrarily weak coupling.  As in one-dimension, we expect that  important quantum corrections will appear primarily in  the energy range 
$E_{\rm{min}} < E < E_0$, where $E_0$ is the nominal bottom of the free-particle band, and $E_B = E_0 - E_{\rm{min}}$ is the binding energy due to self trapping.  Further, we want to have  $E_B > \omega_{\rm{ph}}> T$, for our methods to apply and for quantum corrections to be important.  If the coupling is below the critical value, one obtains $E_B = 0$, so these conditions cannot be satisfied.

Another issue concerns the contribution of short wave length phonons. In three dimensions, the phonon contribution to the particle self-energy has a strong ultra-violet divergence in the continuum limit, so the contribution of short-wavelength phonons may need to be taken into account even in situations where the width of typical particle wave function $\phi(x)$ at the energy of interest is large compared to the lattice constant.  (When quantum fluctuations are taken into account, there is also an ultraviolet divergence in one-dimension, as discussed in  Appendix \ref{prefactors}, but that divergence is logarithmic and is unlikely to be important in practice.) Because the ultraviolet-divergent contribution is only weakly dependent on the particle energy, we may treat it, to a first approximation, as a constant downward shift in the particle energies, which can be taken into account by a (temperature-dependent) redefinition of the  threshold energy $E_0$. A more detailed discussion of how to treat this energy shift may be found in \cite{HL2}, where a similar ultraviolet divergence was encountered in their analysis of the density of states for an  electron in a three-dimensional Gaussian white noise potential.

Additional problems may arise when one tries to extract the density of states $D(E)$ from the imaginary time correlation function $C(\tau)$ obtained from either the ITDH or FWF approximation, using the procedure discussed in Sec. \ref{invert}.  
At least in the classical phonon limit, we know from the Halperin--Lax analysis that at least in limit of  classical phonons and wave functions wide on the scale of the lattice constant,  there will be a region of energy in the low-energy tail where $\ln D(E) \propto -|E-E_0|^{1/2}$, so  $d^2 [\ln D(E ) ]/dE^2 >0 $, which violates the requirements of Sec. \ref{invert}.  This difficulty may be avoided, however, if one adopts a modified procedure, described below, where the Laplace transform is inverted at an earlier stage of the calculation.

In two dimensions, for Gaussian white noise in the continuum limit, the HL analysis predicts $\ln D(E) \propto -|E-E_0|$, so that $d^2 \ln D / dE^2 \approx 0$. This case is marginal, and it is unclear whether one can apply directly the ITDH method to this case. In Ref. \cite{Kim22}, results of a numerical calculation were presented for $D(E)$ of a two-dimensional model of a particle interacting with quasiclassical frozen acoustic phonons. However, that calculation did not extend far enough into the low-energy tail to warrant comparison with an analysis using the methods of the present paper.

\subsubsection{Modified procedures}\label{modifiedaprx}

In the Modified ITDH procedure, one chooses an arbitrary initial phonon configuration $\bm\alpha$  and uses the imaginary-time-dependent Hartree equations to calculate the function $C_{\bm\alpha}(\tau)$ as described Sec. \ref{ITDH}.  Now, however, we use the procedure of Sec. \ref{invert} to find the inverse Laplace transform of $C_{\bm\alpha}(\tau)$ at fixed $\bm\alpha$, which gives the conditional density of states,  
\begin{eqnarray}
D_ { \bm\alpha } (E)  &=&      \sum_{mn}   \braket{ \bm\alpha }{m} \sbra{m}      \psi_{x_0}    \sket{n}    \sbra{n}  \psi^{\dagger}_{x_0}   \sket{ \bm\alpha }
\nonumber \\  & \times &  \delta(E-E_n+E_m) ,
\end{eqnarray}
where the sum is over eigenstates of the Hamiltonian with one particle present for the state $n$ and no particle present for the state $m$. We expect that 
the conditions of Sec. \ref{invert} will be satisfied by   $C_{\bm\alpha}(\tau)$  for fixed  $\bm\alpha $, so there should be no difficulty inverting the Laplace transform at this stage.
We now have
\be
\label{therm1}
D(E)  =  \int   d\bm\alpha     w_{ \bm\alpha }  D_{\bm\alpha} (E) .
\ee
Then, we may approximate the density of states by the value of the integrand in this equation when  $\bm\alpha$ is chosen to maximize its value. [Alternatively, we may approximate the density of states choosing $\bm{\alpha}$ in a way that maximizes
$w_{\bm{\alpha}}C_{\bm{\alpha}}(\tau)$ for the value of $\tau$ that corresponds to the target value of $E$.] 
We expect that in most cases either of these choices will lead to  a good approximation for $D(E)$ in the low-energy tail, and in cases where the averaged correlation function $C(\tau)$ does satisfy the conditions for the procedure of Sec. \ref{invert}, there should be close agreement between the modified ITDH and the original ITDH approximations. It appears that difference between the two procedures will affect only the pre-exponential factors.
As in the previous sections, one should properly include an entropy prefactor $S(E)$, which we have here set equal to 1.

We may proceed in a similar manner when employing the FWF.  
The function $C_{\phi_{\rm{tr}}} (\tau)$ defined by (\ref{CFC}) is the Laplace transform of a function $D_{\phi_{\rm{tr}}} (E) $,  which we may consider to  be an approximation to the actual density of states $D(E)$. One should be able to obtain 
the function  $D_{\phi_{\rm{tr}}} (E)$  from $C_{\phi_{\rm{tr}}} (\tau)$ with good accuracy and without much difficulty,    using the procedure outlined in Sec. \ref{invert}.

Having obtained the estimated density $ D_{\phi_{\rm{tr}}} (E) $ for various choices of the wave function $\phi_{\rm{tr}}$, we choose, for each $E$,  the wave function that maximizes the estimate for that energy.
This defines a modified  FWF approximation:
\be
D(E) = {\rm{max}}_{\phi_{\rm{tr}}} \, [D_{\phi_{\rm{tr}}} (E) ].
\label{maxDOS}
\ee
This should be compared to the FWF approximation of Sec. \ref{FWF}, where we did not compute the functions  $D_{\phi_{tr}}(E)$ but rather obtained the density of states by taking the inverse Laplace transform of the entire function $C(\tau)$. 
 Again, one should get more accurate results if one can include an estimate of the entropy prefactor $S(E)$,  which we have here set equal to 1.

As an alternative to inverting the Laplace transform in the Modified FWF procedure,  if one wishes to obtain a more accurate value for the function $D_{\phi_{\rm{tr}}} (E)$,  one may work directly in the energy regime, following  the prescription of Hopfield \cite{Hopfield62}, which is described in Appendix \ref{HopfieldMethod} below.  This method is not restricted to the low-energy tail, and it can achieve, in principle, an arbitrary degree of accuracy.   However, in the low-energy tail, an approximate inversion of the Lax formula using the method of Sec. \ref{invert} is simpler and probably adequate for the level of approximtaion already implicit in the FWF method.

We remark that there is at least one case where the modified FWF approximation taken literally will lead to nonsensical results. In a model where the lattice vibrations are dispersionless  optical phonons, with a single frequency $\omega_0$, the function $D^0_{\rm{tr}}(E)$ will contain a series of $\delta$-functions at energies separated by multiples of $\omega_0$. Choosing $\phi_{\rm{tr}}$ at each $E$ to maximize  $D^0_{\rm{tr}}(E)$ will simply give an estimated density of states that is infinite at all energies. However, we expect that this pathology will not be a cause for worry when there is at least a  moderate amount of dispersion in the phonon spectrum. 

We expect that in most cases, when $D(E)$ satisfies the condition $d^2 \ln D /dE^2 < 0$,  the difference between the modified procedures and the original ITDH or FWF approximation will not be too great. For the one-dimensional continuum model studied above, we find that the modified FWF result for $D(E)$ only differs from the result of the original FWF approximation by a constant pre-exponential factor of $(3/2)^{1/2}$.  (See Appendix \ref{DOSmax}.)

\section{Summary}
\label{conclusions}

In Sections \ref{invert} to \ref{classical} of this  paper, we  introduced two  related approximation schemes for calculating the momentum-dependent spectral density $A(k,E)$  and the tunneling density of states $D(E)$ in the low-energy tail for the model of a single injected particle coupled to a thermal bath of phonons. In both schemes, we developed an approximation for imaginary-time correlation function $C(\tau)$, which is the Laplace transform of $D(E)$,  and we discussed how $D(E)$ can be efficiently extracted from  $C(\tau)$ for energies $E$ in the  low-energy tail. 
In particular, we obtained a one-to-one relation between energies $E$  and imaginary times $\tau_E$ such that $D(E)$ is determined by $C(\tau)$ and its first two derivatives at $\tau = \tau_E$.

In the ITDH approach, defined in Sec. \ref{ITDH}, we proposed a imaginary-time-dependent Hartree appoximation to calculate the imaginary-time correlation function starting from an initial state
that is in an arbitrary coherent phonon state before the particle is injected. To obtain $C(\tau)$,  one should then calculate the weighted average of this correlation function over a thermal distribution of initial states. However, in the low-energy tail, corresponding to large values of $\tau$,  we can get a good first approximation to $C(\tau) $ considering only the single initial state that gives the largest contribution to the average. In principle, corrections to this approximation can be obtained by using second-order perturbation theory to account for deviations of the initial phonon configuration from the optimal coherent state, as well as corrections arising from the difference between the full Hamiltonian and the Hartree approximation used to calculate the imaginary-time evolution.   

The FWF approximation, introduced in Sec. \ref{FWF}, is a simplification of the ITDH, in which for a given choice of $\tau$, we ignore the time-dependence of the particle portion of the wave function for imaginary times $\tau' < \tau$ by assuming a fixed trial wave function $\phi_{\rm{tr}}(x)$. The phonon configuration is assumed to vary with $\tau' $, however, driven by the coupling to the mean particle density $|\phi_{\rm{tr}}(x)|^2$. Then, for each value of $\tau$, we choose a trial wave function that maximizes the estimate of $C(\tau)$. The FWF approach may be further simplified by restricting the trial wave function to a form controlled by a small number of variational parameters and then choosing the values of these parameters so as to maximize $C(\tau)$.   

As noted in Sec. 
\ref{classical}, in the classical phonon limit, where the nuclear mass is taken to infinity, so that the relevant phonon frequencies are small compared to the temperature $T$, the problems under consideration reduce to calculations of the density of states or the spectral density for a particle in a random potential with a Gaussian statistical distribution.  In the low-energy tail, to leading order (on a logarithmic scale), the ITDH and FWF approximations become equivalent to each other in the classical limit, and their results coincide, at this level, with the results obtained more than five decades ago for a particle in a  random potential. 

In Sec. \ref{continuum}, we presented an application of these methods to a one-dimensional model of a particle interacting via a deformation potential with acoustic phonons in the continuum limit. We presented results of numerical calculations using the FWF approximation for a selected set of parameters, $\gamma, \rho$, and $T$ controlling the particle-phonon coupling strength, the nuclear mass density, and the temperature.   The parameters were chosen such that the phonon frequencies were small compared to the self-trapping energy for the particle in the classical phonon limit, but the ratio between $T$ and  the relevant phonon frequencies could take various values.  Indeed, it is under these conditions that we expect the ITDH and FWF approximations to be most interesting. At least in this parameter region, we found that quantum fluctuations arising from a finite nuclear mass had little effect when the relevant phonon frequencies were small compared to $T$, but tended to increase the density of states at low energies when the phonon frequencies were larger than $T$. In the case where the mass density is held fixed and $T$ is decreased, we found that $D(E)$ remains finite in the limit $T \to 0$ for $E$ greater than $E_{\rm{min}}$, the ground state energy of a self-trapped particle in the classical phonon limit, but $D(E)= 0$ at $T=0$ for $E < E_{\rm{min}}$.
We intend  to present results of an application of the ITDH to the one-dimensional continuum model in a future publication.

The calculations presented in Sec. \ref{continuum} were confined to examples with relatively strong particle-phonon coupling, where it was sensible to consider a case where the self-trapping energy is large compared to the relevant phonon frequencies.  In the opposite case, where the phonon frequencies are large compared to the self-trapping energy, the description may be quite different. In this case, the quantum ground state will be a lightly bound polaron, which is highly mobile with a slightly renormalized effective mass, and its kinetic energy must be taken into account at low temperatures. At high temperatures, the classical phonon approximation can still be made, and the methods for treating a particle in a Gaussian random potential could be used. However, a good description of quantum corrections to the low-temperature low-energy tail for weak particle-phonon interactions requires further investigation.

Several additional topics were discussed in Sec. \ref{additional}. 
In Subsection \ref{AkE} we showed how the momentum-dependent spectral density $A(k,E)$ can be obtained along with $D(E)$ in either the ITDH or FWF approximation. In Subsection \ref{3D} we discussed the modifications that must be made if one wishes to apply either the ITDH or FWF approximation to a three-dimensional system.

In Appendix \ref{EOMITDH} below, we present a detailed derivation of the imaginary-time-dependent Hartree equations of motion  used in the ITDH approximation.
In Appendix \ref{prefactors}, we discuss corrections to the ITDH  that affect the pre-exponential factors in $D(E)$. In particular, we discuss the correction that is of greatest importance  in the case of a continuum system in the classical phonon limit.  This correction arises from fluctuations of the frozen phonon state in which the potential well retains its optimum form but the center of the well is displaced slightly from the position of the injected particle.
Several other topics are treated in additional appendixes. 

Although our study of $A(k,E)$ and $D(E)$ was largely motivated by the problem of optical absorption by an exciton in the presence of a bath of phonons, there are other applications, at least in principle.  As one example, the tunneling density of states measured in an ideal scanning tunneling
microscopy (STM) experiment for an electron injected into an empty band in a two-dimensional insulator should be proportional to the quantity $D(E)$ calculated for that system. Similarly, an inverse angle-resolved photoemission spectroscopy (ARPES) experiment could give a measure of $A(\bfk,E)$.   As a practical matter, however, it is not clear whether one can achieve the sensitivity and energy resolution necessary to probe the low-energy tail region where the methods described above would be directly applicable.  Of course, one must also contend with the influence of impurities in this region, and in the case of an STM measurement, one would have to account for the perturbation caused by the presence of a scanning tip. In addition, it is difficult to perform an STM measurement in a completely empty band, as it is necessary for the target to have at least some lateral conductivity.  

The spectral density $A(\bfk,E)$ or the tunneling density of states $D(E)$ for an occupied electron band can be measured, respectively, by an ARPES or STM experiment.  The problem in this case is that the effects of electron-electron interactions are likely to be larger than the effects of electron-phonon interactions.

A quantity analogous to the spectral density of a particle interacting with phonons in a crystal can arise for an impurity atom injected into an atomic Bose condensate \cite{Shashi2014}. At low energies, the important excitations of the Bose condensate are phonons, and their interaction with the impurity atom may have a form similar to the one considered here, at least in the case of weak coupling.   However, the analogy breaks down when coupling to the impurity is strong \cite{Grusdt2017},  and it is not clear whether one can achieve a regime where a low-energy tail exists and our methods could be directly applicable.

More generally, however, we expect that the type of analysis exemplified by our ITDH procedure may have applicability to a variety of problems where coupling of a particle to degrees of freedom other than phonon modes is important.  For example, excitations about a Fermi sea of electrons may be treated as a set of harmonic-oscillator modes in some circumstances. Also, magnetic excitations in spin systems may often be treated as independent harmonic oscillators.  Appropriate generalizations of the ITDH procedure might be useful for calculations outside the low-energy tail, which would accordingly extend the applicability of the general approach. 

In principle, our methods could be used when the initial phonon state is not a state of thermal equilibrium,  provided it can be described by a density matrix that commutes with the Hamiltonian in the absence of the injected particle and is, therefore,  independent of time.

\section*{Acknowledgments}

The authors are indebted to Eric J. Heller for discussions on possible applications of the coherent state description of phonons to systems of phonons interacting with electrons, which served as a major stimulus for this work.  
  One of the authors (B.I.H.) is pleased to acknowledge his debt to John J. Hopfield, who recommended to him, in 1963, to study a model of excitons interacting with phonons, in an effort to explain the experimentally observed low-energy tail of optical absorption in a variety of insulators.
  We thank the National Science Foundation for supporting this research, through the STC Center for Integrated Quantum Materials (CIQM), NSF  Grant No. DMR-1231319.

\appendix

\section{EQUATIONS OF MOTION FOR IMAGINARY-TIME-DEPENDENT HARTREE APPROXIMATION}\label{EOMITDH}

The unnormalized many-body state $\ket{\Psi(\tau)}$ must satisfy
the  imaginary-time-dependent Schr\"{o}dinger equation
\begin{align}
    \frac{\partial \ket{\Psi}}{\partial \tau}= -H\ket{\Psi}.
    \label{ITDSE}
\end{align}
Assuming that the factorization approximation is valid, the state $\ket{\Psi(\tau)}=a^{-1/2} R(\tau) \Phi_\tau^\dagger \ket{\bm\beta(\tau)}$ should satisfy
\begin{align}
    \braket{\Psi}{\frac{\partial \Psi}{\partial \tau}}
    =
    \braket{\frac{\partial \Psi}{\partial \tau}}{\Psi}
    =
    -\mel{\Psi}{H}{\Psi}
\end{align}
and $\braket{\Psi}=a^{-1}(R(\tau))^2$. Then, one obtains
\begin{align}
    \frac{\partial }{\partial \tau}\ln[(\braket{\Psi})^{-1/2}]
    =
    \frac{\mel{\Psi}{H}{\Psi}}{\braket{\Psi}}
    =
    E_\phi+\sum_k \omega_k \abs{\beta_k}^2,
\end{align}
which gives
\begin{align}
    \frac{d R}{d \tau}
    =
    -R(E_\phi+\sum_k \omega_k \abs{\beta_k}^2).
    \label{ITRE}
\end{align}

The Heisenberg equation of motion for $a_k$ gives
\begin{align}
    da_k/d\tau=-\comm{a_k}{H}=-\omega_k a_k-a\sum_x \psi_x^\dag\psi_x\lambda_k e^{-ikx}.
\end{align}
Then, in the presence of a particle, 
\begin{align}
    d\beta_k/d\tau
    =
    -\omega_k \beta_k-\lambda_k\abs{\phi^2}_k.
\end{align}
Without a particle, 
\begin{align}
    d\beta_{0k}/d\tau=-\omega_k \beta_{0k}.
\end{align}

Multiplying the Schr\"odinger equation \eqref{ITDSE} by $\bra{\bm\beta(\tau)}\psi_x$,  and using the equation \eqref{ITRE} for $d R/d \tau$, one obtains
\begin{align}
    \frac{d\phi(x,\tau)}{d\tau} 
    =
    -&\left[H_e(x)+V(x,\tau)-E_\phi(\tau)-iQ(\tau)\right]\nonumber
    \\
    &\times\phi(x,\tau),
    \label{EOMforphi}
\end{align}
where $H_e(x)$ is a position representation of the electronic kinetic energy operator and 
\begin{eqnarray}
    Q(\tau)
    &=&
    i\bra{\bm\beta(\tau)}\frac{d}{d\tau}\ket{\bm\beta(\tau)} \nonumber \\
    &=&
    -\Im{\sum_k\beta_k^*(\tau)\frac{d\beta_k(\tau)}{d\tau}} \nonumber \\
    &=&
    \Im{\sum_k\beta_k^*(\tau)\lambda_k\abs{\phi^2}_k},
\end{eqnarray}
where we have used  $\ket{\beta_k(\tau)}=e^{-\abs{\beta_k(\tau)}^2/2}e^{\beta_k(\tau)a_k^\dag}\ket{0}_k$
and Eq. \eqref{EOMbeta}.

Note that $\lambda_{-k}=\lambda_{k}^*$ due to time-reversal symmetry and $\abs{\phi^2}_{-k}=\abs{\phi^2}_{k}^*$ since $\phi(x)^2$ is real.
If we choose an initial configuration such that $\alpha_{-k}=\alpha_{k}^*$, then $\beta_{-k}(\tau)=\beta_{k}^*(\tau)$ and $\beta_{0,-k}(\tau)=\beta_{0k}^*(\tau)$ since the equations of motion preserves the relations.
Then, the purely imaginary term vanishes, i.e., $Q(\tau)=0$, meaning that the propagated electronic wave function $\phi(x_0,\tau)$ is real if the initial wave function $\phi(x_0,0)$ is real. Furthermore, the inner product $\prod_k\sbraket {\beta_{0k}(\tau)} {\beta_k(\tau)}$ is real for this initial configuration. Thus, $C_{\bm\alpha}(\tau)$ is real for the initial configuration. 

If we choose an initial configuration such that
$C_{\bm\alpha}(\tau)$ is complex, there will be a complex-conjugate initial configuration, with the same weight $w_{\bm\alpha}$, which will give rise to the complex-conjugate value of $C_{\bm\alpha}(\tau)$.
Therefore, if there is a unique 
initial configuration that maximizes $w_{\bm\alpha}\abs{C_{\bm\alpha}(\tau)}$, it must be real, with an initial  condition that satisfies $\alpha_{-k}=\alpha_{k}^*$.  We believe that this will generally be the case.

We expect that the imaginary-time-dependent Hartree approximation should become asymptotically exact in the limit where $T$ and the phonon frequencies $(\propto M^{-1/2})$ go to zero.  In this limit, evolution of the phonon coordinates is slow, and the particle wave function will adiabatically follow the ground state wave function for a  particle in the potential well produced by the phonon configuration.  This is the limit where a Born-Oppenheimer separation is valid, and corrections to the Hartree approximation become small, despite the fact that the imaginary times $\tau$ of interest grow proportional to $T^{-1}$ or $M^{1/2}$. (Note that there is no Fock exchange term in our problem, since we have only one mobile particle.)

\section {HIGHER-ORDER CORRECTIONS} \label{prefactors}

In order to improve our estimates of the density of states in the low-energy tail, we need to examine the pre-exponential factors $S(\tau)$,  introduced in  earlier sections, which we have so far ignored.  
As mentioned previously,  the most important corrections to the ITDH approximation can be calculated, in principle, by treating the difference between the actual Hamiltonian and the imaginary-time-dependent Hartree approximation and the deviation of the actual starting configuration $\bm\alpha$ from the optimum configuration as small perturbations, whose effects one can estimate using second-order time-dependent perturbation theory.  We will not attempt to implement this procedure in the present paper. 

In the classical phonon limit, the analysis is simplified, because for long times $\tau$, for a given potential configuration $V(x)$, the correlation function $C_V(\tau)$  is determined by properties of the electronic ground state in this potential. Then, the corrections to the ITDH estimation of $C(\tau)$ can be calculated using time-independent perturbation theory to account for  deviations of the actual potential $V(x)$ from its optimum form.  For a continuum model, it turns out that the most important correction arises from fluctuations where the potential well retains its ideal form but where the center of the well is displaced slightly from the position $x_0$ where the particle is injected. 

To be more precise, let us assume that  for a given value of $\tau$, the optimum  potential has a form
\be
V_{\rm{opt}}(x) = u (x-x_0) 
\ee
where the shape of $u$ is independent of $x_0$, and let us write the ground state wave function as
\be
\phi_B(x)  = f(x-x_0) .
\ee   
The ground state in a potential well can always be chosen to be a real-valued function with no nodes, and it is necessarily non-degenerate. Moreover,  we expect by symmetry that the optimum potential will have a minimum at $x=x_0$ and will be symmetric about that point, so that the ground state wave function in the well will have a maximum at $x_0$. 

Let $C_0(\tau)$ be the estimate of $C(\tau)$ obtained from (\ref{Cmax})  with the optimum choice of $V$. 
We may now estimate  the contribution to $ C(\tau)$ from a displaced potential of the form $V(x) = u(x - x_0 -s)$, for $s \neq 0$.  The displacement $s$ will have no effect on the weight factor $w_{\bm\alpha}$ in Eq. (55) nor on the binding energy $E_B$, but it will reduce the value of $|\phi_B(x_0)|^2$  by a factor $|f(s)/f(0)|^2$. We may obtain an improved estimate of $C(\tau)$ by integrating over the displacement $s$, namely
\be
C_1(\tau) = S_1(\tau) C_0(\tau) ,
\ee
\be
S_1(\tau) = \nu_0 \int ds |f(s) / f(0)|^2 ,
\ee
where $\nu_0$ is the density per unit length of independent choices of $s$ 
 (Here, we have assumed that the  bound state wave function is broad on the scale of the lattice constant $a$, so we have taken the continuum limit $a \to 0$, and we have replaced the sum over positions by an integral).
 
We may determine $\nu_0$ as follows.  Consider a set of potentials of the form  
\be
V_\eta (x) = u(x-x_0 )  - \eta u' (x-x_0) ,
\ee
 with a parameter $\eta$.   
 Since  the probability of $V_\eta$ is controlled by the weight function $w_{\bm\alpha}$, the variable $\eta$ will have a Gaussian distribution of the form
\be
p(\eta) = (2 \pi \sigma_\eta^2)^{-1/2} e ^{- \eta^2 / 2 \sigma_\eta^2} ,
\ee
with 
\begin{align}
    \sigma_\eta^{-2}
    &=
    \int dx dx'   u'(x) G(x-x')  u'(x')
    \\
    &=
    \int_{- \pi /a} ^{\pi / a}
    \frac{dk}{2\pi}\frac{ \rho \omega_k  } {|\gamma_k|^2 (n_k+1/2)}
    k^2\abs{\tilde{u}(k)}^2,
    \label{sigmaetam2}
\end{align}
where as was defined in Eq. \eqref{Kdef},
\be
G(x-x') =  \int_{- \pi /a} ^{\pi / a} \frac{dk}{2 \pi  }  \frac{ \rho \omega_k  } {|\gamma_k|^2 n_k} e^{i k (x-x')} ,
\label{Gqc}
\ee
and $\tilde{u} (k) = \int dx e^{-ikx} u(x) .$
But a small nonzero value of $\eta$ is equivalent to a displacement of the potential by an amount $s = \eta$, so we must have 
\be
\nu_0 = \lim_{\eta \to 0} p( \eta) = (2 \pi \sigma_\eta^2)^{-1/2} .
\ee
Since the wave function $f(s)$ is normalized to unity, we obtain
\be
S_1(\tau) = (2 \pi \sigma_\eta^2 |f(0)|^4  )^{-1/2}  .
\label{S1originaleq}
\ee

 The corrected correlation function $C_1$ leads to a density of states  similar to that obtained in Ref. \cite{HL1} using  a minimum counting procedure, which approximated $D(E)$ by the density of local minima of the smoothed potential $V_s$ with 
$V_s(x)  = \theta - E $.  The factor $S_1$ represents the correction  
imposed by the requirement that $V_s$ is a local minimum at a point $x$, on top of the requirement that the value of $V_s(x)$ is equal to $\theta-E$.

For the acoustic phonon model discussed in Sec. \ref{continuum},
\begin{align}
    u(x)=-\frac{\kappa^2}{m}\sech^2(\kappa x),
\end{align}
\begin{align}
    \tilde{u}(k)=-
    \frac{k\pi}{m}\csch(\frac{k\pi}{2\kappa}),
\end{align}
and $\gamma_k=\gamma\abs{k}$.
For classical phonon, $n_k\approx T/\omega\gg1$, so Eq. \eqref{sigmaetam2} gives
\begin{align}
    \sigma_\eta^{-2}
    \approx
    \frac{ K  } {\gamma^2 T}
    \int_{- \infty} ^{\infty}
    \frac{dk}{2\pi}
    k^2\abs{\tilde{u}(k)}^2
    =
    \frac{ K  } {\gamma^2 T}
    \frac{16\kappa^5}{15m^2},
\end{align}
where similar approximations were used as in Eq. \eqref{classicalFtau}.
Then, in Eq. \eqref{S1originaleq} using $f(0)=(\kappa/2)^{1/2}$ [c.f. Eq. \eqref{phitr}], one finds
\begin{align}
    S_1 = \left( \frac {32K\kappa^3} {15\pi m^2 \gamma^2T} \right)^{1/2} .
    \label{S1factor}
\end{align}

Although the correction $S_1$ was derived for the ITDH approximation, it seems reasonable to apply it also to the FWF approximation.  Doing this, one obtains Eq. \eqref{DOSclassicalS1}, which only differs from the exact asymptotic value \eqref{DOSGWNAS} by a constant factor of $(2/15)^{1/2}$.
If one applies  the correction to the modified FWF approximation,
then, from \eqref{S1factor} and \eqref{DOSclassicalrhomax}, one obtains
\begin{align}
    S_1(E) \, D_{MFWF}(E)
    =
    \frac{1}{\sqrt{5}}
    \frac{4\kappa^2}{\pi m \xi}
    e^{-\frac{4\kappa^3}{3m^2\xi}},
\end{align}
which differs from the exact asymptotic value \eqref{DOSGWNAS} by a factor of $1/\sqrt{5}$ and is the same as the density of states in Gaussian white noise obtained in  Eq. (4.9) of Ref. \cite{HL1} using the approximtion that counted the minima of the smoothed potential. 

Contributions to the pre-exponential factor beyond those included in $S_1$ come from fluctuations in $V(x)$ that are orthogonal to $u(x-x_0)$ and $u'(x-x_0)$.  As discussed in \cite{HL2}, for the Gaussian white noise potential, these corrections lead to a finite numerical correction $S_2$ to the density of states in the low-energy tail, which is independent of the energy or the strength of the disorder potential.  Moreover, a calculation based on second order perturbation theory is sufficient to obtain  results which coincide with the exact asymptotic form of the density of states  (\ref{DOSGWNAS}) \cite{ZittartzLanger}. 

When quantum fluctuations are taken into account, corrections arising from short-wavelength phonons need to be handled with additional care.  As mentioned above, these fluctuations lead to an ultraviolet divergence in the self-energy in the  continuum model, even in one dimension.  Specifically, in second order perturbation theory, one obtains a self-energy
\begin{align}
    \Sigma (k, E) \approx -  \int dk \frac { (2n_k+1)  |\gamma_k|^2 } { 4 \pi \rho \omega_k (\epsilon_k - E)}  .
\end{align}
In the classical phonon limit, where $2n_k+1 \to 2T/\omega_k$, the integral converges at large $k$. When quantum fluctuations are included, however, the integral has a logarithmic divergence at large $k$, giving a contribution to the self-energy of form 
$( m\gamma^2/ \pi \rho c_s) \ln a$, where $c_s\propto \rho^{-1/2}$  is the sound velocity and $a$ is the short-distance cutoff. 

In situations where the resulting self-energy is large, it may be most  convenient to treat the contribution from  short-wavelength fluctuations as a downward shift of the bare energy spectrum $\epsilon_k$, while including the remaining fluctuations in a calculation of the pre-expoential factors in the density of states.

\section{POTENTIAL FLUCTUATIONS AND GAUSSIAN WHITE NOISE}

From Eq. \eqref{Hep}, one obtains the operator-valued potential
\be
\hat{V}(x)=\sum_k (  \lambda_k a_k^\dagger + \lambda_{-k}^* a_{-k} ) e^{-ikx} ,
\ee
and the result
\begin{align*}
    \ev{\hat{V}(x)\hat{V}(x+\delta x)}_{x}
    &\equiv
    L^{-1}\int dx \hat{V}(x)\hat{V}(x+\delta x)
    \\
    &=
    \sum_{k} \frac{\abs{\gamma_k}^2}{2\rho L \omega_k}
    (a_k+a_{-k}^\dag)(a_{k}^\dag+a_{-k})e^{-ik\delta x}.
\end{align*}
Then, the average of the correlation function over the thermal ensemble of phonons at temperature $T$ is
\begin{align}
    \ev{\hat{V}(x)\hat{V}(x')}_{T}
    &=
    \sum_{k} \frac{\abs{\gamma_k}^2}{2\rho L \omega_k}
    (2n_k+1)e^{-ik(x-x')}
    \\
    &=
    \int_{-\pi/a}^{\pi/a} \frac{dk}{2\pi} \frac{\abs{\gamma_k}^2}{2\rho \omega_k}
    (2n_k+1)e^{-ik(x-x')}.
    \label{AC}
\end{align}
If one uses the quasiclassical potential
\be
V_{\rm{qc}}(x)=\bra{\bm\alpha}\hat{V}(x)\ket{\bm\alpha}=\sum_k (  \lambda_k \alpha_k^* + \lambda_{-k}^* \alpha_{-k} ) e^{-ikx} ,
\label{Vqc}
\ee
its spatial autocorrelation function averaged over the thermal ensemble of phonons is 
\begin{align}
    \ev{V_{\rm{qc}}(x)V_{\rm{qc}}(x')}_{T}
    &=
    \sum_{k} \frac{\abs{\gamma_k}^2}{\rho L \omega_k}
    n_ke^{-ik(x-x')}
    \\
    &=
    \int_{-\pi/a}^{\pi/a} \frac{dk}{2\pi} \frac{\abs{\gamma_k}^2}{\rho \omega_k}
    n_ke^{-ik(x-x')},
    \label{ACqc}
\end{align}
missing the quantum fluctuation contribution to Eq. \eqref{AC} \cite{Kim22}.
In the classical phonon limit, $\rho\to\infty$ while fixing $T$,
\begin{align*}
    n_k=\frac{1}{e^{\omega_k/T}-1}\approx\frac{T}{\omega_k}\propto \sqrt{\rho}\gg1,
\end{align*}
which gives
\begin{align}
    \ev{\hat{V}(x)\hat{V}(x')}_{T}
    \approx
    \int_{-\pi/a}^{\pi/a} \frac{dk}{2\pi} \frac{\abs{\gamma_k}^2T}{\rho \omega_k^2}
     e^{-ik(x-x')}.
     \label{ACcl}
\end{align}

\subsection{Continuum acoustic phonon model}
For the continuum acoustic phonon model in Sec. \ref{continuum}, $\gamma_k=\gamma\abs{k}$ and $\omega_k$ given by Eq. \eqref{dispersion}, the autocorrelation in the classical phonon limit [Eq. \eqref{ACcl}] becomes
\begin{align*}
    \ev{\hat{V}(x)\hat{V}(x')}_{T}^{\rm{cont}}
    \approx
    \frac{\gamma^2T}{K}\int_{-\pi/a}^{\pi/a}\frac{dk}{2\pi} 
    \frac{(ka/2)^2}{\sin^2(ka/2)}e^{-ik(x-x')}.
\end{align*}
Then, its integration gives
\begin{align}
    \int_{-\infty}^{\infty}d(x-x') \ev{\hat{V}(x)\hat{V}(x')}_{T}^{\rm{cont}}
    =
    \frac{\gamma^2T}{K}.
    \label{ACinteg}
\end{align}
\subsection{Gaussian white noise}
For Gaussian white noise, the potential spatial autocorrelation is given by
\begin{align*}
        \ev{V(x)V(x')}=\frac{\xi}{2}\delta(x-x')
\end{align*}
Then, its integration gives
\begin{align}
    \int_{-\infty}^{\infty}d(x-x') \ev{V(x)V(x')}
    =
    \frac{\xi}{2}
    \label{GWNinteg}
\end{align}
By comparing Eqs. \eqref{ACinteg} and \eqref{GWNinteg}, one obtains the relation between the parameters in the Gaussian white noise model and the continuum model,
\begin{align}
    \frac{\xi}{2}=\frac{\gamma^2T}{K}.
    \label{GWNcontinuumParams}
\end{align}

\subsection{Gaussian statistics}\label{qcweight}
The Fourier component of the quasiclassical potential \eqref{Vqc} is
\begin{align}
    \tilde{V}(k)=\int dx e^{-ikx}V_{\rm{qc}}(x)
    =
    L\lambda_{k}^*(\alpha_{k}+\alpha_{-k}^*).
    \label{Vtildek}
\end{align}
Thus, if the function $V(x)$ is specified, the combinations $\alpha_k + \alpha^*_{-k}$ are thereby determined, but no information is gained about the quantities  $ \eta_k = \alpha_k - \alpha^*_{-k}$.
Note that
\begin{align}
    \sum_k \frac{\abs{\alpha_k}^2}{n_k}
    &=
    \sum_k \frac{\abs{\alpha_k+\alpha_{-k}^*}^2+\abs{\alpha_k-\alpha_{-k}^*}^2}{4n_k}
    \\
    &=
    \sum_k
    \left[\frac{\tilde{G}(k)\abs{\tilde{V}(k)}^2}{2L}
    +
    \frac{\abs{\eta_k}^2}{4n_k}
    \right],
\end{align}
where we made use of Eqs. \eqref{Gqc} and \eqref{Vtildek}, and $\tilde{G}(k)=\int dx e^{-ikx}G(x)$.
Then, $w_{\bm\alpha}$ [C.f. Eq. \eqref{walpha}] can be decomposed into the product of functions of $\tilde{V}(k)$ and $\eta_k$, meaning $\tilde{V}(k)$ and $\eta_k$ are independent random variables.
If we integrate $w_{\bm{\alpha}} $ over the variables $\eta_k$, we obtain a probability distribution for the function $V(x)$:
\begin{align}
    w_{\rm{qc}}
    &=
    Z_{\rm{qc}}^{-1}
    \exp[-\sum_k
    \frac{\tilde{G}(k)\abs{\tilde{V}(k)}^2}{2L}]
    \\
    &=
    Z_{\rm{qc}}^{-1}
    \exp[-\frac{a^2}{2}\sum_{x,x'}
    V(x) G(x-x') V(x')],
\end{align}
where  $Z_{\rm{qc}}$ is a normalization constant.

\section{MODIFIED FWF APPROXIMATION}\label{DOSmax}
If we apply the Modified FWF approximation to the one-dimensional Gaussian white noise potential, 
 the imaginary-time correlation function $C_{\phi_\kappa}(\tau)$ can be analytically continued to the real-time correlation function through the relation $\tau=it$:
\begin{align*}
    \tilde{C}_{\phi_\kappa}(t)=C_{\phi_\kappa}(it)= \abs{\phi_\kappa(x_0)}^2e^{-it E_e^{\phi_\kappa}-\sigma_{\phi_\kappa}^2 t^2/2}.
\end{align*}
Since this is a simple Gaussian form, we can easily take the Fourier transform 
to obtain the corresponding estimate of the density of state
[c.f. Eq. \eqref{corrFourier}]:
\begin{align}
    \label{Dphikappa}
    D_{\phi_\kappa}(E) = 
    \frac{\kappa}{2}e^{-3(E- \kappa^2/6m)^2/\xi\kappa}\left(\frac{3}{\pi\xi\kappa}\right)^{1/2}.
\end{align}
We get precisely the same result if we use Laplace's method to obtain the inverse Laplace transform of $C_{\phi_\kappa}(\tau)$ directly [C.f. Eq. \eqref{laplace}].
Note that Eq. \eqref{EdlnCdt} predicts $E
    =
    E_e^{\phi_\kappa}-\sigma_{\phi_\kappa}^2\tau_E$,
giving a linear relation between $\tau_E$ and $E$: $\tau_E=(E_e^{\phi_\kappa}-E)/\sigma_{\phi_\kappa}^2$.

Now the density of states  $D_{\phi_\kappa}(E)$ can be maximized with respect to the variational parameter $\kappa$. The value of $\kappa$ that maximizes the density of states is found from
\begin{align}
    \frac{d\ln D_{\phi_\kappa}(E)}{d\kappa}
    =
    \frac{d}{d\kappa}
    \left[
    \ln\kappa
    -
    \frac{1}{2}\ln \sigma_{\phi_\kappa}^2
    -\frac{(E-E_e^{\phi_\kappa})^2}{2\sigma_{\phi_\kappa}^2}\right]
    =0.
    \label{kappamaxeqrho}
\end{align}
Since this equation is second order in $E$, it can be solved for $E<0$:
\begin{align}
    E=-\frac{\kappa^2}{6m}-\sqrt{\left(\frac{\kappa^2}{3m}\right)^2-\frac{\gamma^2 T}{3K}\kappa}.
\end{align}
Inverting this equation for $\kappa$ gives the optimal $\kappa_{\rm{o}}(E)$ that maximizes the density of states for the given energy $E$, from which we obtain
\begin{align}
    D_{\rm{MFWF}}(E)
    \equiv
    {\rm{max}}_\kappa[D_{\phi_\kappa}(E)]
    =
    D_{\kappa_{\rm{o}}(E)}(E).
    \label{rho1cont}
\end{align}
In the low-energy limit $E\to-\infty$, the optimal $\kappa_{\rm{o}}(E)$ reduces to the Halperin--Lax result $\kappa_{\rm{o}}(E)\to\kappa_E=(-2mE)^{1/2}$ \cite{HL1}.
Then, the asymptotic form of $D_{\rm{MFWF}}$ can be obtained:
\begin{align}
    D_{\rm{MFWF}}(E)
    \approx
    \left(\frac{3\kappa_E}{4\pi \xi}\right)^{1/2}e^{-\frac{4\kappa_E^3}{3m^2\xi}}&\text{ for }E\ll0.
    \label{DOSclassicalrhomax}
\end{align}
This result differs from $D_{\rm{FWF}}(E)$, given by Eq. (\ref{DOSclassical}),   by a factor of $(3/2)^{1/2}$.

\section{HOPFIELD'S METHOD FOR TREATING THE FRANCK-CONDON PROBLEM}\label{HopfieldMethod}

An application of Hopfield's method to our problem  proceeds by introducing  a function $D^\eta_{\phi_{\rm{tr}}}   (E)$ which is equal to the trial density of states $D_{\phi_{\rm{tr}}} (E)$ for a problem where all the coupling constants $C_k$ are multiplied by a constant $\eta^{1/2}$, with $0 \leq \eta \leq 1$. Following Hopfield's arguments,  $D_{\phi_{\rm{tr}}}$ may be obtained by solving the ``transport equation''
\be
\frac {\partial D^\eta_{\phi_{\rm{tr}}} (E)}{\partial \eta} = \int d E'  \tilde{K} (E-E') D^\eta_{\phi_{\rm{tr}}} (E') ,
\ee
with the kernel
\begin{eqnarray}
\tilde{K} (\epsilon) & = &  
\sum_k \frac {|C_k|^2 }{ \omega_k^2 } [ (n_k +1) \delta (\epsilon - \omega_k)    \nonumber   \\     &+&  n_k \delta (\epsilon + \omega_k ) + 
( \omega _k \frac{\partial} {\partial \epsilon}  - 2n_k  -1 ) \delta(\epsilon)  ]  , 
\end{eqnarray} 
and the initial condition 
\be
D^0_{\phi_{\rm{tr}}} (E) = |\phi_{\rm{tr}} (x_0) |^2 \delta (E-E^\phi_e) .
\ee
One then identifies $ D_{\phi_{\rm{tr}}} (E) $ with $ D^1_{\phi_{\rm{tr}}} (E)  $.

\bibliography{refs}

\end{document}